%% file: ch3oh_final_for_arXiv.tex
\documentclass[aps,jcp,superscriptaddress,amsmath,amssymb]{revtex4-1}

\usepackage[unicode]{hyperref}
\usepackage{amsmath,amssymb} 
\usepackage{float} 
\usepackage[T1]{fontenc}  
\usepackage{multirow}
\usepackage{dcolumn} 
\usepackage[flushleft]{threeparttable}
\usepackage{booktabs}
\usepackage{color,soul}

\hypersetup{
   unicode=true,          
   plainpages=false,
   colorlinks=true,       
   linkcolor=blue,          
   citecolor=blue,        
}
\usepackage{url}

\usepackage{graphicx} 
\usepackage{amsmath}
\usepackage{amstext}
\usepackage{amsfonts}
\usepackage{amssymb}
\usepackage{color}
\usepackage{longtable}
\usepackage{threeparttable}
\usepackage{rotating}
\usepackage{multirow}
\usepackage{tikz}
\usepackage[version=4]{mhchem}
\usepackage{hyperref} 
\usepackage{listings}
\usepackage{float}
\usepackage{threeparttable}
\usepackage{url}
\usepackage{hyperref}
\usepackage{here}
\usepackage[absolute,overlay]{textpos}

\newcommand{\be}{\begin{equation}}
\newcommand{\ee}{\end{equation}}
\newcommand{\bea}{\begin{eqnarray}}
\newcommand{\eea}{\end{eqnarray}}

\newcommand{\eqa}{\begin{equation}}
\newcommand{\eqz}{\end{equation}}
\newcommand{\eqma}{\begin{eqnarray}}
\newcommand{\eqmz}{\end{eqnarray}}

\newcommand{\dd}{\text{d}}

\newcommand{\bos}[1]{\boldsymbol{#1}}

\usepackage{array}
\newcolumntype{R}[1]{>{\raggedleft  \arraybackslash}p{#1}@{} }
\newcolumntype{C}[1]{>{\centering \arraybackslash}p{#1}@{} }

\newcommand{\cm}{$\text{cm}^{-1}$}

\newcommand{\ctau}{$\xi(\tau)$ }
\newcommand{\xitau}{$\xi(\tau)$}
\newcommand{\Ltau}{$L(\tau)$}
\newcommand{\Cs}{$C_\mathrm{s}$}
\newcommand{\Ctv}{$C_{\mathrm{3v}}\mathrm{(M)}$}
\newsavebox{\mycode} %
\newcommand{\stext}{\text{s}}
\newcommand{\tvec}{$\bos{t}$}
\newcommand{\detg}{\tilde{g}}
\newcommand{\iim}{\text{i}}
\newcommand{\tref}{{\text{ref}}}

\usepackage[unicode]{hyperref}
\usepackage{soul}
\hypersetup{
   unicode=true,          
   plainpages=false,
   colorlinks=true,       
   linkcolor=black,          
   linkcolor=blue,          
   citecolor=blue,        
   urlcolor=blue           
}
\usepackage{url}

\begin{document}

\title{Variational Vibrational States of Methanol (12D)}
\author{Ayaki Sunaga}
\author{Gustavo Avila}%
\author{Edit M\'atyus}%
\email{edit.matyus@ttk.elte.hu}
\affiliation{%
ELTE, E\"otv\"os Lor\'and University, Institute of Chemistry, P\'azm\'any P\'eter s\'et\'any 1/A 1117 Budapest, Hungary
}%

\date{\today}

\begin{abstract}
\noindent %
    Full-dimensional (12D) vibrational states of the methanol molecule (\ce{CH3OH}) have been computed using the GENIUSH-Smolyak approach and the potential energy surface from Qu and Bowman (2013). 
    All vibrational energies are converged better than 0.5~\cm\ with respect to the basis and grid size up to the first overtone of the CO stretch, ca. 2000 \cm\ beyond the zero-point vibrational energy.
    About seventy torsion-vibration states are reported and assigned. 
    The computed vibrational energies agree with the available experimental data within less than a few \cm\ in most cases, which confirms the good accuracy of the potential energy surface.    
    The computations are carried out using curvilinear normal coordinates with the option of path-following coefficients which minimize the coupling of the small- and large-amplitude motions. 
    It is important to ensure tight numerical fulfilment of the $C_{3\text{v}}$(M) molecular symmetry for every 
    geometry and coefficient set used to define the curvilinear normal coordinates along the torsional coordinate to obtain a faithful description of degeneracy in this floppy system.
    The reported values may provide a computational reference for fundamental spectroscopy, 
    astrochemistry, and for the search of the proton-to-electron mass ratio variation using the methanol molecule.  \\
\end{abstract}

\maketitle

%
%
\section{Introduction}
Understanding the large-amplitude motions (LAM) of floppy molecules and complexes has been a fundamental problem in the quantum dynamics of nuclear motion. 
The traditional and commonly used rigid rotor and harmonic oscillator approximations \cite{Born1925ZP_harm, Huber1979_mol_spectra, Papousek1982_mol_vib_tra, Gordy_MMS} have often been employed for systems consisting of only small-amplitude motions (SAMs). However, these models and Hamiltonians for the SAMs cannot describe the energy levels contributed by the LAMs even qualitatively, and theory development has been required \cite{Ito1987JPC_LAM,Sorensen1979_LAM,Laane2012_non-rigid,Matyus2023CC}. LAMs do not appear only in artificially made molecular complexes; many organic and biological molecules include one or several LAMs in the methyl unit (\ce{-CH3}) and the hydroxyl unit in CXYZ-OH type compounds. Small systems, with a single LAM, include HCOOH recently studied in Refs.~\citenum{Daria2022JMS_HCOOH,Avila2023PCCP_HCOOH} and \ce{CH3OH}, which is the subject of the present study.

The methanol molecule has been a prototype for torsion-rotation coupling in high-resolution spectroscopy experiments \cite{Lees1968JCP_torrot_exp,Gerry1976JMS_torrot_exp,Herbst1984JMS_torrot_exp,Xu1996JMS_torrot_exp} and theoretical modelling \cite{Hougen1994JMS, Mekhtiev1996JMS_torrot_theo, Xu1999JCP_torrot_theo, Xu2002JMS_torrot_theo}, 
and the hyperfine-torsional coupling has recently been also investigated \cite{Belov2016JCP_hyperfine,Lankhaar2016JCP_hyperfine,Lankhaar2018NA_hyperfine,Xu2019JMS_hyperfine}.
Nevertheless, the accurate computation of the torsional-vibrational levels (in full dimensionality), is still challenging. The effective torsional Hamiltonians have been used in simulating the high-resolution spectrum, and the parameters of the model Hamiltonian have been fitted to experimental results \cite{Wang1998JCP_fit,Hanninen1999JCP_fit}. We also mention the use of 1-dimensional (1D) reduced-dimensionality models \cite{Caro1997CPL,Blasco2003CPL_ch3oh} in this context. These modelling efforts were followed by full-dimensional \emph{ab initio} computations using \emph{ab initio} potential energy surfaces in conjunction with the internal-coordinate-path Hamiltonian (ICPH) \cite{Tew2003MP_ICPH} and the reaction-path Hamiltonian \cite{Fehrensen2003JCP, Bowman2007JPCA} used in variational vibrational computations, as well as a vibrational perturbation theory approach with internal-coordinate quartic force fields \cite{Sibert2005JCP}. Recently, quantum dynamics computations of the methanol molecule in full dimensionality have been reported \cite{Lauvergnat2014SA_methanol,Nauts2018MP_methanol}. Despite earlier efforts, even the most accurate computation so far \cite{Nauts2018MP_methanol} has a significant deviation from experiment: more than 5~\cm\ deviation for the vibrational band origin (VBO) at 353.2 \cm\ and 510.3 \cm, and more than 9~\cm\ for the VBOs at 751.0 \cm\ and 1046.7 \cm.

Bowman and co-workers have developed two \emph{ab initio,} full-dimensional potential energy surfaces (PESs) for \ce{CH3OH}, 
invariant to all permutations of identical nuclei. The 2007 version (PES07) was obtained at the coupled cluster singles, doubles,
and non-iterative triples correction (CCSD(T)) level of theory \cite{Bowman2007JPCA}, while the 2013 version (PES13) was extended to the \ce{CH3OH -> CH3 + OH} dissociation regime and was fitted to electronic energies obtained at the CCSD(T)-F12b level for the single-reference region and the second‐order perturbation theory based on a complete active space (CASPT2) for the multi-reference region \cite{Qu2013MP_CH3+OH}. The PES07 has been employed in full-dimensional (12D) variational vibrational computations \cite{Lauvergnat2014SA_methanol,Nauts2018MP_methanol}. 
Most recently, the zero-point vibrational energy of methanol and deuterated methanol was reported for PES13.\cite{Nandi2019JCC}

\ce{CH3OH} has attracted interest in areas beyond fundamental spectroscopy. 
The variation of the constants such as the $\mu$ proton-to-electron mass ratio as a function of time and/or position \cite{Yahil1975_delta_mu,Murphy2001MNRS_var_cons,Uzan2002RMP_var_cons,Martins2017RPP_var_cons} may be detected in the molecular spectra. It is known that the fractional change in the $\nu$ vibrational frequency is proportional to the fractional variation of the proton-to-electron mass ratio, $\Delta \mu/\mu \approx \kappa_\nu \Delta\nu/\nu$.
One of the directions includes using (ultra-coolable) diatomic molecules \cite{Flambaum2007PRL_var_cons,DeMille2008PRL_delta_me,Zelevinsky2008PRL_delta_mu}, but \ce{CH3OH} and isotopologues offered a promising alternative due to the sensitivity of $\Delta \mu/\mu$ enhanced in near-degenerate torsional energy levels and torsion-rotation coupling \cite{Jansen2011PRL_ch3oh,Levshakov2011APJ_methanol,Kanekar2015MNRSL_ch3oh,Vorotyntseva2023MNRAS_ch3oh}. Methanol also plays an important role in astrochemistry.
Since methanol has been observed in a variety of astronomical environments in large quantities, its formation path \cite{Fuchs2009AA_form,Wirstrom2011AASS_ch3oh_ast_reac,Nakai2023APJ_ch3oh_ast_reac} and reaction \cite{Olah2016JACS_ch3oh,Ishibashi2023APJ} have attracted attention. 
Understanding the quantum dynamics of isolated methanol itself is essential for investigations in complex astronomical environments including several molecules and molecular complexes. On the other hand, as summarized in our study 
(Sec. \ref{subsec:compare_exp}),
some of the coupled torsion-vibrational states have never been reported neither in laboratory experiments, nor in computations.

Methanol has twelve vibrational degrees of freedom, including one large-amplitude torsional motion, and its full-dimensional variational computation had been challenging. 
Recent and ongoing developments have made it possible to perform variational vibrational computations for
9D \cite{Viglaska2022JMS_H2O-HF,Felker2021JPCA_HCl_H2O,Vindel-Zandbergen2023JCP_H2O-HCN}, 12D \cite{Lauvergnat2014SA_methanol,Nauts2018MP_methanol,Wang2018JCP_12D,
Avila2019JCP_CH4F-,Avila2019JCP_methodology,Simmons2023JCP_new_collocation,Avila2020PCCP,Papp2023MP,Felker2023_HF3,Wang2023JCP_H2O_dim,Simko2024JCP}, and 15D \cite{Kallullathil2023JCP_contraction}
molecular systems with large-amplitude motions.
In this context, we mention the full, 21D-dimensional variational vibrational computation of malonaldehyde \cite{Lauvergnat2023CPC_malon}. Another direction towards larger system-sizes in variational quantum dynamics approaches relies on a system-bath separation \cite{Chen2022JCTC_bath}.

The present paper reports the computation of all vibrational states up to ca.~2000~\cm\ beyond the zero-point vibrational energy (ZPVE) of CH$_3$OH converged within 0.5~\cm. The vibrational states are assigned based on the analysis of the wave function. The computed and experimental VBOs agree within a few \cm\ in most cases where experimental data is available.
The rest of the paper is organized as follows. 
Section~\ref{sec:theory} introduces the vibrational coordinate definition and implementation details regarding the torsion-following curvilinear normal coordinates and the numerical representation of the $C_{3\text{v}}(\text{M})$ molecular symmetry (MS) group in the computations. 
Section~\ref{sec:metho} defines the basis and grid pruning conditions. 
Section~\ref{sec:results} reports 
the computed vibrational energies, their assignment and comparison to the available experimental values. 
Section \ref{sec:conclusion} is summary, conclusion, and outlook.

%
%
\section{Theory and Implementation}\label{sec:theory}
\subsection{Classical and quantum mechanics of molecular rotations and vibrations \label{sec:clqrovib}}
To introduce all necessary notation, we start this introduction with the classical Lagrangian of $N$ atomic nuclei experiencing the $V$ potential energy,
\be
\label{eq:lagrangian}
  L 
  = 
  \frac{1}{2} \sum_{i=1}^N m_i \dot{\bos{X}}_i^{\mathrm{T}} \dot{\bos{X}}_i - V \; .
\ee
For an efficient description of molecular motions, the laboratory-fixed Cartesian coordinates, $\bos{X}_i$ are replaced
by the $\bos{X}^{\mathrm{NCM}}$ translational Cartesian coordinates of the nuclear centre of mass (NCM), 
the $\Omega=(w_1,w_2,w_3)$ rotational coordinates, and the corresponding $\bos{r}_i$ body-fixed Cartesian coordinates,
according to
\be
  \bos{X}_i
  =
  \bos{O}(\Omega)\bos{r}_i 
  +
  \bos{X}^{\mathrm{NCM}}.
\ee
Furthermore, bond distance, angle, and torsion-type internal coordinates, $\bos{\rho}=(\rho_1, \rho_2, \ldots, \rho_{3N-6})$, are defined as a function of $\bos{r}_i\ (i=1,\ldots,N)$. 
As a short notation for the new coordinates, we introduce
\be
\boldsymbol{\zeta} = \{\zeta_1, \zeta_2, \ldots, \zeta_{3N}\} 
= 
\{\rho_1, \rho_2, \ldots, \rho_{3N-6}, \omega_1, \omega_2, \omega_3, X^{\mathrm{NCM}}_X, X^{\mathrm{NCM}}_Y, X^{\mathrm{NCM}}_Z\}.
\ee
The Lagrangian for this new set of coordinates is obtained with the chain rule as
\be
  L=\frac{1}{2} \sum_{k l=1}^{3N}  \dot{\zeta}_k g_{kl} \dot{\zeta}_l-V, 
\ee
where
\be\label{eq:g_t}
  g_{kl} 
  = 
  \sum_{i=1}^N m_i 
    \frac{\partial \bos{X}_i^{\mathrm{T}}}{\partial \zeta_k} \frac{\partial \bos{X}_i}{\partial \zeta_l} \; .
\ee
is the mass-weighted metric tensor. 
The elements of $\bos{g}$ can also be written with \tvec-vectors, \emph{e.g.,} Ref.~\cite{Matyus2023CC},
\be\label{eq:small_g}
  g_{kl} = \sum_{i=1}^N m_i \boldsymbol{t}^{\mathrm{T}}_{ik}\boldsymbol{t}_{il} ,
\ee
where the vibrational $t$-vector is
\be\label{eq:vib_t}
  \boldsymbol{t}_{i k}
  = 
  \frac{\partial \boldsymbol{r}_i}{\partial \rho_k} \; , \quad k=1,2, \ldots, 3N-6 \; ,
\ee
and the rotational $t$-vector is
\be\label{eq:rot_t}
  \boldsymbol{t}_{i 3N-6+a}
  =
  \boldsymbol{e}_a \times \boldsymbol{r}_i\; , \quad a=1(x),2(y), 3(z) 
\ee
with the $\boldsymbol{e}_a$ unit vector pointing along the body-fixed axis $a$.

The $p_k$ generalized momentum canonically conjugate to the general $\zeta_i$ coordinate is
\be
  p_k
  =
  \frac{\partial L}{\partial \dot{\zeta}_k}
  =
  \sum_{l=1}^{3N} g_{k l} \dot{\zeta}_l 
  \quad\rightarrow\quad 
  \boldsymbol{\dot{\zeta}}
  =
  \boldsymbol{g}^{-1}\boldsymbol{p} \; .
\ee
By defining 
\begin{align}
  \bos{G}=\bos{g}^{-1} \; ,
  \label{eq:Ginvg}
\end{align}
we have an expression with the momenta, 
\be
  H
  =
  \frac{1}{2} 
  \sum_{k=1}^{3N}
  \sum_{l=1}^{3N}
    p_k G_{kl} p_l + V \; .
\ee
The translation-rotation and translation-vibration blocks of $\bos{g}$ and $\bos{G}$ are identically zero.  Henceforth, $\bos{g}$ and $\bos{G}$ collects the $k,l=1,\ldots,3N-3$ elements corresponding to the rotational and vibrational degrees of freedom (after translation is separated exactly).

The quantum Hamiltonian is obtained through the correspondence principle with considering also Podolsky's note~\cite{Podolsky1928PR} regarding quantization in curvilinear coordinates, 
\begin{align}
  \hat{H}^\text{rv,P}
  =&
  \frac{1}{2}
  \sum_{k=1}^{3N-3} \sum_{l=1}^{3N-3}
    \detg^{-1/4} \hat{p}_k \detg^{1/2} G_{kl} \hat{p}_l  \detg^{-1/4} + V
    \label{eq:hampod}
\end{align}
where $\detg=\det\bos{g}$, the generalized momentum operators are $\hat{p}_{k}=-\iim\hbar\partial / \partial \rho_k \ (k=1,\ldots,3N-6)$,
and $\hat{p}_{3N-6+a}=-\iim\hbar\partial / \partial \omega_a=\hat{J}_a$ correspond to 
the components of the body-fixed angular momentum.

The present work focuses on vibrational computations, so we will continue with the vibration-only part (corresponding to $J=0$ rotational angular momentum states),
\begin{align}
  \hat{H}^\text{v,P}
  =
  \frac{1}{2}
  \sum_{k=1}^{3N-6} \sum_{l=1}^{3N-6}
    \detg^{-1/4} \hat{p}_k \detg^{1/2} G_{kl} \hat{p}_l  \detg^{-1/4}
  +V \; .
  \label{eq:hampodvib} 
\end{align}
A computationally efficient, full-dimensional solution of the vibrational Schrödinger equation requires basis and grid truncation (Sec.~\ref{sec:metho}), which was most conveniently implemented for the `fully rearranged' form of the operator,  
\cite{Avila2019JCP_methodology,Avila2019JCP_CH4F-,Avila2020PCCP,Daria2022JMS_HCOOH,Avila2023PCCP_HCOOH,Papp2023MP,Matyus2023CC}
\be
  \hat{H}^{\mathrm{v}}
  =
  -\frac{\hbar^2}{2} \sum_{k=1}^{3N-6} \sum_{l=1}^{3N-6} G_{kl} \frac{\partial}{\partial \rho_k} \frac{\partial}{\partial \rho_l}
  -\frac{\hbar^2}{2} \sum_{k=1}^{3N-6} B_k \frac{\partial}{\partial \rho_k}
  +U
  +V,
  \label{eq:Hfrearr}
\ee
with
\begin{align}
  B_k
  &=
  \sum_{l=1}^{3 N-6} \frac{\partial}{\partial \rho_l} G_{l k} \;, 
  \label{eq:Bkin}
  \\ 
  U
  &=
  \frac{\hbar^2}{32} 
  \sum_{k l=1}^{3 N-6}
  \left[%
    \frac{G_{k l}}{\tilde{g}^2} 
    \frac{\partial \tilde{g}}{\partial \rho_k} \frac{\partial \tilde{g}}{\partial \rho_l}
    +
    4 \frac{\partial}{\partial \rho_k}\left(\frac{G_{k l}}{\tilde{g}} \frac{\partial \tilde{g}}{\partial \rho_l}\right)
  \right] \; .
  \label{eq:Ukin}
\end{align}
This presentation corresponds to the GENIUSH implementation of the numerical kinetic energy approach \cite{Matyus2009JCP} (and a recent review is in Ref.~\citenum{Matyus2023CC})
For pioneer computer implementations of the numerical kinetic energy operator approach, we mention the TNUM approach \cite{Lauvergnat2002JCP_TNUM,Lauvergnat2014SA_methanol}
(the TANA approach \cite{Ndong2012JCP_TANA}) implementations by Lauvergnat and co-workers, and the TROVE approach \cite{Yurchenko2007JMS_TROVE,Yurchenko2017JCTC_sym-TROVE}
originally based on the s-vector formalism \cite{Sorensen1979_LAM}, by Yurchenko and co-workers.

\subsection{Primitive coordinates for CH$_3$OH vibrational computations \label{sec:primitiveInternal}}

\begin{figure}[ht]
\includegraphics[width=0.5\textwidth]{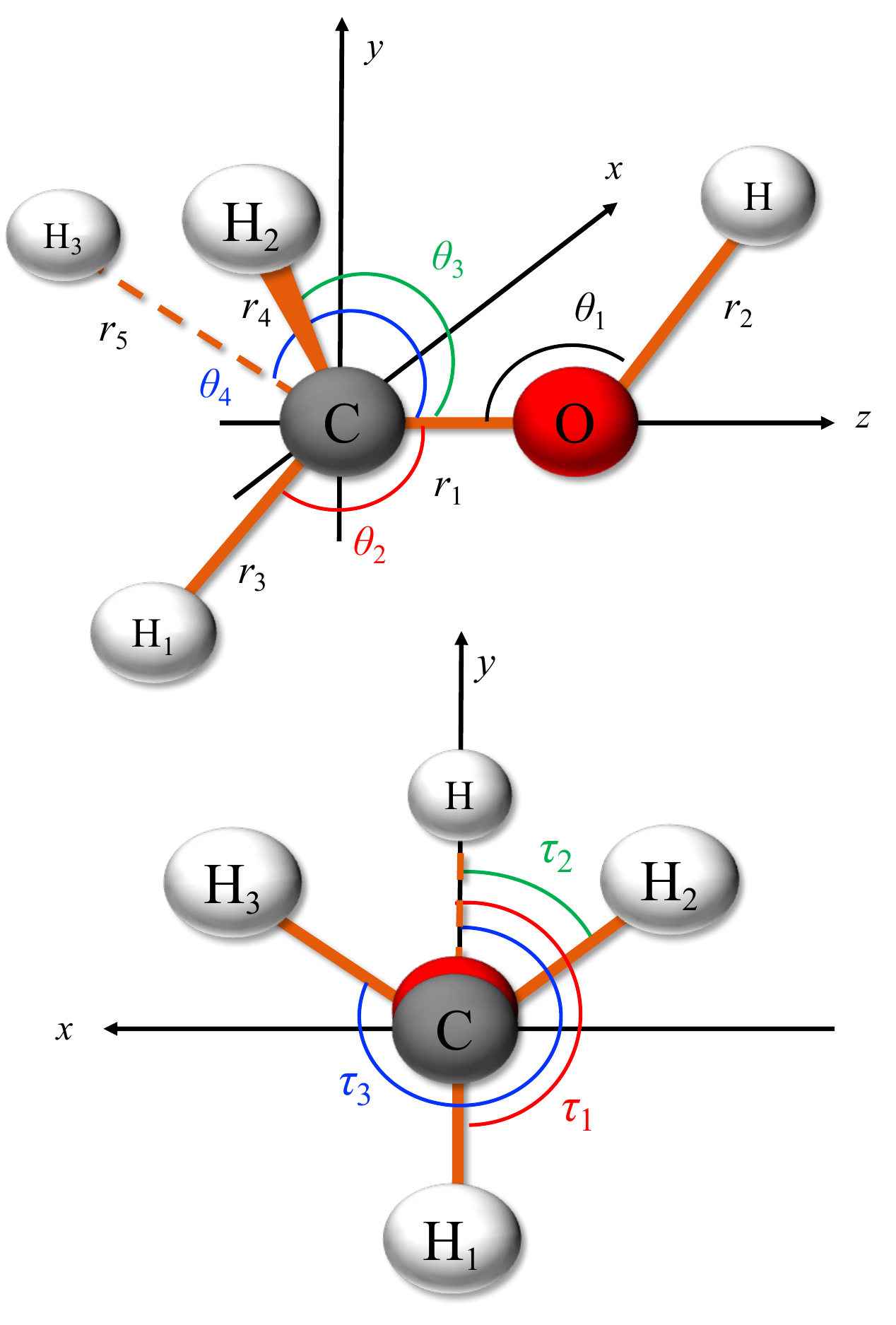}
\caption{%
  Visualization of the methanol molecule and the primitive internal coordinates in the right-handed body-fixed Cartesian frame. 
  \label{fig:methanol_xyz_sym} 
} 
\end{figure}

For a physically motivated coordinate representation of molecular motions, it is often a good choice 
to use distance- ($r_i$), angle- ($\theta_i$), and dihedral-angle-type ($\tau_i$) curvilinear internal coordinates.
A particular choice of the internal coordinates and the body-fixed frame can be defined by specifying the body-fixed Cartesian coordinates in terms of the selected internal coordinates. 
For the methanol molecule studied in this work, we use the following coordinate definition (see also Table \ref{tbl:zmat_methanol_sym} and Figure \ref{fig:methanol_xyz_sym}),
\bea\label{eq:cart_sym}
&\boldsymbol{r}_{\mathrm{C}}=  \mathbf{0}, \quad \boldsymbol{r}_{\mathrm{O}}=\left(\begin{array}{c}0 \\ 0 \\ r_1\end{array}\right), \\ \nonumber
&\boldsymbol{r}_{\mathrm{H}}=  \boldsymbol{r}_{\mathrm{O}}+\left(\begin{array}{c}0 \\ r_2 \cos \left(\theta_1-\pi / 2\right) \\ r_2 \sin \left(\theta_1-\pi / 2\right)\end{array}\right), \\ \nonumber
&\boldsymbol{r}_{\mathrm{H}_1}=  \left(\begin{array}{c}-r_3 \sin \theta_2 \sin \tau_1 \\ r_3 \sin \theta_2 \cos \tau_1 \\ r_3 \cos \theta_2\end{array}\right), \\ \nonumber
&\boldsymbol{r}_{\mathrm{H}_2}=  \left(\begin{array}{c}-r_4 \sin \theta_3 \sin \tau_2 \\ r_4 \sin \theta_3 \cos \tau_2 \\ r_4 \cos \theta_3\end{array}\right), \\ \nonumber
&\boldsymbol{r}_{\mathrm{H}_3}=  \left(\begin{array}{c}-r_5 \sin \theta_4 \sin \tau_3 \\ r_5 \sin \theta_4 \cos \tau_3 \\ r_5 \cos \theta_4\end{array}\right) .
\eea
Furthermore, the centre of mass is fixed at the origin of the body-fixed coordinate system, which amounts to an additional constant shift vector to each vector in Eq.~\eqref{eq:cart_sym}.
Throughout this work, we use atomic masses for the nuclei, $m_\text{O}= 15.994915$~u, $m_\text{C}=12$~u, $m_\text{H}=1.007825$~u, which is a common choice in polyatomic rovibrational computations,\cite{Kutzelnigg2007MP_nuc_atom_mass} (in spite of the fact that the Born--Oppenheimer (BO) approximation prescribes nuclear masses for the nuclear motion, which can be corrected at higher perturbative orders \cite{Matyus2018JCP_mass_correc,Matyus2019JCP_non_adia}). 
The frame fixed to the non-rigid body is also defined by the body-fixed Cartesian coordinate definition, Eq.~\eqref{eq:cart_sym}. In this vibrational study, we continue using the definitions in Eq.~\eqref{eq:cart_sym} (without prescribing any further rotation of the frame), but we plan to implement the Eckart--Sayvetz condition \cite{Eckart1935PR,Sayvetz1939JCP,Papousek1982_mol_vib_tra} or a path-following Eckart frame \cite{Lauvergnat2016JCP} in future work, for efficient rovibrational computations. 

Instead of the $\tau_1,\tau_2$, and $\tau_3$ coordinates, Eq.~\eqref{eq:cart_sym}, we use their linear combination, following earlier work of Sibert and co-worker \cite{Sibert2005JCP} and Lauvergnat and co-workers \cite{Blasco2003CPL_ch3oh,Lauvergnat2014SA_methanol,Nauts2018MP_methanol},
\bea \label{eq:lin_com}
 \quad \tau&=& \frac{1}{3}\left(\tau_1+\tau_2+\tau_3\right), \\ \nonumber
 \quad\varphi_{1}&=&\frac{1} {\sqrt{2}}\left(\tau_2-\tau_3\right), \\ \nonumber
 \quad\varphi_{2}&=&\frac{1} {\sqrt{6}}\left(2 \tau_1-\tau_2-\tau_3\right) \; ,
\eea
which allows us to identify $\tau$ as the single LAM in the system, with the $\varphi_1,\varphi_2$ SAMs,
 and to account for the \Ctv\ molecular symmetry. 

For later convenience, we define the following compact notation for the internal coordinates,
\be
  \boldsymbol{\rho} 
  =
  \left(\boldsymbol{\xi}; \tau \right)=\left(r_1, r_2, r_3, r_4, r_5, \theta_1, \theta_2, \theta_3, \theta_4, \varphi_1, \varphi_2; \tau\right) ,
  \label{eq:rhoxitau}
\ee
where $\boldsymbol{\xi}$ collects the SAMs. The mathematically allowed coordinate ranges are defined as $r_i \in[0, \infty)$ for distances, the $\theta_i \in[0, \pi]$ for angles and $\tau,\tau_i \in[0, 2\pi)$ for the torsional angles. 
To obtain the coordinate range of $\varphi_1$ and $\varphi_2$, we assume the difference between $\tau_i$ and $\tau_j$  ($i\ne j$) is exactly $2\pi/3$ at the equilibrium geometry, and the overlap of two hydrogens does not occur ($\tau_2 < \tau_1 < \tau_3$, where $\tau_2 \in[-2\pi/3, 4\pi/3]$, $\tau_1 \in[0, 2\pi]$, and $\tau_3 \in[2\pi/3, 8\pi/3]$). 
Under this condition, the mathematically allowed ranges for the dihedral angles are $\varphi_1 \in(-2\pi/\sqrt{2}, 0)$ and $\varphi_2 \in(-2\pi/\sqrt{6}, 2\pi/\sqrt{6})$, respectively, and the equilibrium geometries are  $\varphi^{\mathrm{eq}}_1 = -4\pi/(3\sqrt{2})$ and $\varphi^{\mathrm{eq}}_2 =0$. 

%
\begin{table}
\caption{%
  Internal coordinates of methanol. 
  H is the hydrogen atom of the OH unit, H$_i\ (i=1,2,3)$ label the hydrogens of the CH$_3$ unit. 
  The equilibrium values (for one of the three equilibrium structures) for PES13 \cite{Qu2013MP_CH3+OH} are in square brackets, in {\AA} for the distances and in degree for the angles. 
  } \label{tbl:zmat_methanol_sym}
\begin{tabular}{@{}ccc cc cl@{}}
\hline\hline
$\mathrm{C}$ & & & & & &  \\
O & C & $r_1$ $[1.422]$ & \\
 $\mathrm{H}$ & $\mathrm{O}$ & $r_2$ [0.959] &  $\mathrm{C}$ & $\theta_1$ [108.6] \\
 $\mathrm{H}_1$ & $\mathrm{C}$&  $r_3$ [1.088] & $\mathrm{O}$&  $\theta_2$ [106.8] & $\mathrm{H}$ & $\tau_1$ [180.0] \\
 $\mathrm{H}_2$ & $\mathrm{C}$&  $r_4$ [1.095] & $\mathrm{O}$ & $\theta_3$ [112.0] & $\mathrm{H}$ & $\tau_2$ [61.5] \\
 $\mathrm{H}_3$ & $\mathrm{C}$ & $r_5$ [1.095] & $\mathrm{O}$ & $\theta_4$ [112.0] & $\mathrm{H}$ & $\tau_3$ [298.5] \\
\hline\hline
\end{tabular}
\end{table}

%
%

%
%
\subsection{Curvilinear normal coordinates}\label{subsec:G_and_F}
The GF method and the curvilinear (c-normal) coordinates are well known
\cite{Wilson1955book,Papousek1982_mol_vib_tra,McCoy1991JCP_curvilinear,Castro2017MP_c-normal,Perveaux2017PCCP_c-normal,Lauvergnat2014SA_methanol,Nauts2018MP_methanol,Lauvergnat2023CPC_malon,Daria2022JMS_HCOOH,Avila2023PCCP_HCOOH}, 
but we briefly reiterate details to introduce the notation necessary for further discussions. 

The core idea is to define a linear combination of the primitive internal coordinates (Sec.~\ref{sec:primitiveInternal}) that reduce coupling both in the kinetic and in the potential energy (within certain approximations) in the neighbourhood of a reference structure. 
The relevant procedure is known for small-amplitude vibrations about a well-defined reference structure. So, we define the displacement from the reference structure of the small-amplitude coordinates,
\begin{align}
  \Delta \xi_i 
  = 
  \xi_i - \xi_i^\tref \; , \quad i=1,\ldots,D_\text{s}
  \label{eq:dxidef}
\end{align}
where $D_\stext$ is the number of small-amplitude coordinates (assuming to be the first $D_\stext$ coordinates among the internal coordinates).
We aim to define the $Q_1,\ldots,Q_{D_\text{s}}$ coordinates as a(n invertible) linear transformation of the `old' displacement coordinates, 
\bea\label{eq:L_Q}
  \Delta \xi_i = \sum_{k=1}^{D_\stext} L^Q_{ik} Q_k,   
\eea
so that the coupling among the new $Q_k$ coordinates is small. 
The conditions for the smallness of the coupling about the reference structure had been straightforwardly formulated, \emph{e.g.,} Ref.~\citenum{Papousek1982_mol_vib_tra}, for the classical kinetic and potential energy (in the harmonic approximation and for the small-amplitude coordinates, labelled with `s'), 
\begin{align}
  2T
  &=
  \dot{\bos{\xi}}^{\mathrm{T}}(\bos{G}^\stext)^{-1}\dot{\bos{\xi}} 
  =
  \dot{\bos{Q}}^{\mathrm{T}}\bos{L}_Q^{\mathrm{T}}(\bos{G}^\stext)^{-1}\bos{L}_Q\dot{\bos{Q}} 
  =
  \dot{\bos{Q}}^{\mathrm{T}}\bos{E}\dot{\bos{Q}} \;, 
  \label{eq:TclassQ}
  \\
  2V
  &=
  {\bos{\xi}}^{\mathrm{T}}(\bos{F}^\stext) {\bos{\xi}} 
  =  
  \bos{Q}^{\mathrm{T}}\bos{L}_Q^{\mathrm{T}}{\bos{F}^\stext}\bos{L}_Q\bos{Q}
  =
  \bos{Q}^{\mathrm{T}}\bos{\Lambda} \bos{Q} \;,
  \label{eq:VclassQ}  
\end{align}
where we aim to choose $\bos{L}_Q$ so that both the $\bos{E}$ (kinetic energy) and the $\bos{\Lambda}$ (potential energy) matrices are diagonal. Thereby, the new small-amplitude coordinates are uncoupled (at this harmonic level of description).
This requirement is fulfilled by the diagonalization of the real, non-symmetric `GF' matrix, which implies  $\bos{L}^{\mathrm{T}}_Q\neq\bos{L}^{-1}_Q$, 
\bea\label{eq:GFL}
  \bos{L}^{-1}_Q{\bos{G}^\stext}{\bos{F}^\stext}\bos{L}_Q 
  =
  \bos{\Lambda} \; .
\eea
So, first, we solve the eigenproblem (by numerical methods)
\be\label{eq:GFA}
  \textbf{A}^{-1}{\bos{G}^\stext}{\bos{F}^\stext}\textbf{A} 
  =
  \mathbf{\Lambda} \; ,
\ee
and then, fix the norm of the eigenvectors to fulfill Eqs.~\eqref{eq:TclassQ}--\eqref{eq:VclassQ}, according to
\be\label{eq:normalization}
  L^Q_{ik} 
  =
  A_{ik}{N_{k}} \;, 
  \quad\quad  
  N_{k}
  =
  \left(\frac{\lambda_{k}}{\sum_{i,j=1}^{D_\stext} F^\stext_{ij}A_{ik}A_{jk}}\right)^{\frac{1}{2}}.
\ee
$\lambda_{k}$ is the $k^{\mathrm{th}}$ diagonal element of $\boldsymbol{\Lambda}$, and it is related to the harmonic frequency $\nu_k$ through $\lambda_{k}=4\pi^2\nu_k^2$.
The $Q_k\ (k=1,\ldots,D_\stext)$ coordinates defined through Eq.~\eqref{eq:L_Q} with coefficients computed through this GF procedure are called mass-weighted curvilinear normal coordinates. 
In computations, it is useful to work with dimensionless c-normal coordinates, $\bos{q}$,  defined as
\be\label{eq:dimless_q}
  q_k
  =
  \left(\frac{\lambda_k}{\hbar^2}\right)^{\frac{1}{4}} Q_k \; , 
\ee
and therefore, Eq.~\eqref{eq:L_Q} can be rewritten as
\begin{align}
  \Delta \xi_i 
  &=
  \sum_{k=1}^{D_\stext} 
    L^Q_{ik} 
    \left(\frac{ \hbar^2}{\lambda_k}\right)^{\frac{1}{4}} q_k 
  \nonumber \\ 
  &= 
  \sum_{k=1}^{D_\stext} 
    L_{ik} q_k 
  \label{eq:xi_Lq}  
\end{align}
with
\begin{align}
  L_{ik} 
  = 
  L^Q_{ik} 
  \left(%
    \frac{\hbar^2}{\lambda_k}
  \right)^{\frac{1}{4}} \; .
  \label{eq:lcoeff}
\end{align}
This conversion simplifies the computer implementation because $\nu_k$ does not appear in the integration (Appendix~III of Ref.~\citenum{Wilson1955book} and Appendix~E of Ref.~\citenum{Papousek1982_mol_vib_tra}).

To compute the linear combination coefficients, Eq.~\eqref{eq:lcoeff}, for the c-normal coordinates, Eq.~\eqref{eq:xi_Lq}, we first compute the small-amplitude vibrational block of the $\bos{G}$ matrix, Eqs.~\eqref{eq:g_t}--\eqref{eq:Ginvg},
\be
    G^{\stext}_{kl} = G_{kl}(\bos{\rho}^\tref) \; , \quad k,l=1,2,...,D_\stext \; ,
\ee
using the vibrational and rotational \tvec-vectors evaluated at $\bos{\rho}^\tref$ ($i=1,\ldots,N$),
\be
\label{eq:vib_t}
  \boldsymbol{t}_{i k}
  = 
  \left.\frac{\partial \boldsymbol{r}_i}{\partial \rho_k} \right|_{\bos{\rho}=\bos{\rho}^{\mathrm{ref}}}, \; k=1,2, \ldots, 3N-6
  \quad\quad\text{and}\quad\quad
  \boldsymbol{t}_{i ,3N-6+a}
  =
  \boldsymbol{e}_a \times \boldsymbol{r}^{\mathrm{ref}}_i, \; a=1(x),2(y), 3(z) \; .
\ee
Then, the Hessian of the $V$ potential energy is computed at $\bos{\rho}^\tref$ corresponding to the small-amplitude coordinates, 
\be
\label{eq:F_ij}
  \left.F^s_{ij}
  =
  \frac{\partial^2 V(\boldsymbol{\xi})}{\partial \xi_i\partial \xi_j}\right|_{\bos{\rho}=\bos{\rho}^{\mathrm{ref}}}\;,
  \quad i,j=1,2,...,D_\stext \; .
\ee
The Hessian is computed by numerical differentiation (in general, analytic first and second derivatives are not available for PES subroutines) by a specialized central numerical differentiation algorithm (implemented during this work), which 
allowed us to minimize the numerical noise with double-precision arithmetic.
In the outline of the algorithm (Figure~\ref{fig:fortran_code}), the coefficient (\texttt{coef(1:nffd)}) for each step (\texttt{step(1:nffd)}) and corresponding denominator (\texttt{denom}) is obtained in the \texttt{step\_coef} subroutine, where \texttt{nffd} is the order of the numerical differentiation. For instance, when \texttt{nffd = $4$}, \texttt{step(1:4) = [$-2,-1,1,2$]$\times$delta}, \texttt{coef(1:4) = [$1,-8,8,-1$]}, and \texttt{denom = $-144$}. 
\begin{figure*}[ht]
\lstset{%
  }
\begin{lstlisting}[language=fortran,
xleftmargin=0.2\textwidth,
        xrightmargin=0.2\textwidth]
call step_coef(step,coef,delta,denom,nffd)
mdelta = -delta

do j = 1,nSAM
  do i = j,nSAM
   if(i.ne.j)then
     do k = 1,nffd
        do l = 1,nffd
          qnDIM(i) = qref(i) + step(k)
          qnDIM(j) = qref(j) + step(l)
          call potcall(qnDIM,energy)
          Fs(i,j) = energy*coef(k)*coef(l) + Fs(i,j)        
        enddo
     enddo
   else                                                                                                                                                                                                  
     do k = 1,nffd
        do l = 1,nffd
          qnDIM(i) = qref(i) + step(k) + step(l)
          call potcall(qnDIM,energy)
          Fs(i,j) = energy*coef(k)*coef(l) + Fs(i,j)       
        enddo
     enddo
   endif
   Fs(i,j) = Fs(i,j)/(denom*delta*delta)
   Fs(j,i) = Fs(i,j)
   qnDIM(i) = qref(i)
   qnDIM(j) = qref(j)
  enddo
enddo	
\end{lstlisting}
\caption{Outline of the algorithm used for the generalized numerical differentiation. \texttt{nSAM} is $D_\text{s}$, \texttt{qnDIM} stores the internal coordinate values, \texttt{qref} is for the reference coordinates, \texttt{potcall} is the subroutine which returns the potential energy for \texttt{qnDIM}, \texttt{delta} is the step size. 
}
\label{fig:fortran_code}
\end{figure*}

The outlined approach is implemented in the GENIUSH program package for an arbitrary $N$ number of atoms, $3N-7\leq D_\text{s}\leq 3N-6$ number of small-amplitude vibrational coordinates, and $V$ potential energy surfaces (assumed to be available as a linkable 
subroutine or function, indicated as \texttt{potcall} in the figure). We plan to expand the code for several LAMs in future work.

The selection of the $\bos{\rho}^\tref=(\bos{\xi}^\tref,\tau^\tref)$ `reference' structure requires further discussion. 
Regarding the present vibrational computations for CH$_3$OH, the simplest choice for the reference structure is one of the three equilibrium structures of the methanol molecule (Table~\ref{tbl:zmat_methanol_sym} specifies one of them, the other two can be generated by simple symmetry considerations). The couplings of the small-amplitude motions can be minimized near the reference structure, which is expected to allow efficient truncation of the direct-product basis and grid (Secs.~\ref{subsec:plune_basis}--\ref{sec:grid}), since the small-amplitude configuration space relevant for the quantum dynamics is near the reference (equilibrium) structure. At the same time, the coordinate range of the large-amplitude motion relevant for the quantum dynamics is not limited to a small neighbourhood of equilibrium structure(s), LAMs are delocalized over multiple PES wells. For this reason, we have excluded the LAM from the c-normal coordinate definition, Eq.~\eqref{eq:xi_Lq}. If the SAMs-LAM coupling is non-negligible, then SAMs, which may be optimal near one of the equilibrium structures, may be less good near another PES well. To account for the change of couplings among SAMs along the LAM(s), we introduce averaged and LAM-following c-normal coordinates in the following subsections.

\subsubsection{LAM-following curvilinear normal coordinates}\label{subsubsec:LAM-following}

For the case of methanol, we consider a $\tau$-dependent reference structure for the SAMs, $\bos{\xi}^\tref(\tau)$, define displacement coordinates with respect to this reference value for every $\tau$, and perform the GF analysis, Eqs.~\eqref{eq:dxidef}--\eqref{eq:F_ij}, as a function of the $\tau$ coordinate,
\bea
  \xi_i(\tau) 
  &=& 
  \xi_i ^{\mathrm{ref}}(\tau) + \Delta \xi_i(\tau) 
  \; , 
  \label{eq:xitau}
  \\ 
  \Delta \xi_i(\tau) 
  &=&   
  \sum_{k=1}^{D_\stext} L_{ik}(\tau) q_k \; ,
  \quad i=1,\ldots,D_\stext \; .
  \label{eq:dxitau}
\eea
We define $\xi^{\mathrm{ref}}_i(\tau)$ as the geometry which minimizes $V$ for the given $\tau$ value. Henceforth, we call it `relaxed geometry', for short. The $\bos{G}^\stext$ and $\bos{F}^\stext$ matrices are computed according to these choices.
%

%
%
Although curvilinear normal coordinates have been used previously in GENIUSH-Smolyak computations for the example of the formic acid molecule \cite{Daria2022JMS_HCOOH,Avila2023PCCP_HCOOH}, the coordinate coefficients were computed (on a case-by-case basis) using Wolfram Mathematica. 
The present work reports the general implementation of the (path-following) GF method and curvilinear normal coordinates in the GENIUSH program (in Fortran).
In our implementation, the path-following GF method is automatically performed for the optimized geometry at each $\tau_n$ point ($\bos{\xi}^{\mathrm{ref}}(\tau_n)$), and the coefficient vectors are interpolated, as explained later. The computational time of the curvilinear coordinates along the relaxed path is negligible, and the present implementation points towards the general applicability of the approach.

Regarding the technical details and the numerical precision of the \Ltau\ coefficients, the $\bos{G}^\stext$ matrix is obtained with good numerical stability by numerical differentiation from the GENIUSH main program \cite{Daria2021JCP} at the selected $\bos{\rho}^\tref$ geometry. 
At the same time, the simple two-point central difference formula did not provide a sufficiently precise $\bos{F}^\stext$ matrix. 
To overcome this difficulty, it was necessary to implement the generalized numerical differentiation scheme in Fortran (Figure~\ref{fig:fortran_code}).
The precision of the newly implemented GF code was extensively tested by comparison with the tailor-made Mathematica version (Sec.~\ref{subsec:res:Ltau}), and as a result, it provides a generally applicable, robust implementation within GENIUSH for `any molecule' with 1 (or 0) LAM. Extension to multiple LAMs is possible, but further tests and developments will be required for the multi-dimensional interpolation and the related phase adjustment (\emph{vide infra}). 

Coordinates that are optimal for some $\tau_n$ torsional coordinate value are not necessarily optimal for different $\tau$ values. So, we repeated the computation
over an equidistant grid of $\tau$ points in the $[0,360]^\text{o}$ interval, \emph{i.e.,}
$\bos{\xi}^\tref(\tau_n)$ and $\bos{L}(\tau_n)$ were computed at $\tau_n=(n-1)6^\text{o}\ (n=1,\ldots,61)$. 
(The redundancy at $0^\text{o}$ and $360^\text{o}$ was included for test purposes and the phase adjustment of $\bos{L}(\tau_n)$).
The dataset was interpolated by cubic spline. 
In the present implementation, the derivatives at the boundaries (0 and 360$^\circ$) are obtained by expanding the region, \emph{e.g.,} by substituting the value at $\tau = 354^\circ$ for the one at $\tau = -6^\circ$, we can obtain the first and second derivatives at $\tau = 0, 360^\circ$. 

In the next paragraph, we discuss details of the geometry relaxation, which is followed by details regarding the $\bos{L}$ linear combination coefficients.

\vspace{0.5cm}
\paragraph{$\bos{\xi}^\tref$ relaxation at $\tau_n$ points}
For a selected $\tau_n$ value, the $V(\tau_n)$ 
cut of the PES is minimized. 
This minimum search is carried out in two steps. 
First, the numerical gradient of $V(\tau_n)$ with respect to the SAMs was used to (roughly) locate the minimum, and then, the $\bos{\xi}^\tref$ geometry at the minimum was tightly located using the simplex method \cite{Numerical_recipes_fortran90}. 
With this procedure, we confirmed the minimum as it was reported in Ref.~\citenum{Qu2013MP_CH3+OH}, and pinpointed the minimum energy at a ca.~$8.0 \times 10^{-6}$~\cm\  lower value than the energy reported by the PES developers. 
Such small differences are, of course, most often irrelevant in practical applications. During the definition of the c-normal coordinates, we had to pay attention to tiny numerical differences, and for this reason, it was necessary to implement the high-precision (restricted) minimum-search algorithm just described. We observed that without this careful numerical procedure, numerical imperfections (noise) in the c-normal coordinate coefficients caused small splittings in degenerate energies. These small splittings were convergence errors for the given coordinate definitions, which were expected to disappear in the infinite basis set (and exact integration) limit, but slowed down the convergence of finite-basis computations.

So, for an efficient computational approach, we must have c-normal coordinates which faithfully represent the molecular symmetry of the system. We paid attention to computing the relaxed $\bos{\xi}^\tref(\tau_n)$ values to high numerical precision, \emph{i.e.,} 
(i) the $C_\text{s}$ point-group symmetry at the three equivalent minima and saddle points, 
and 
(ii) the periodicity for $\tau$ (from 0 to 360 degrees). 

\begin{figure*}[!htbp]
\includegraphics[width=0.8\textwidth]{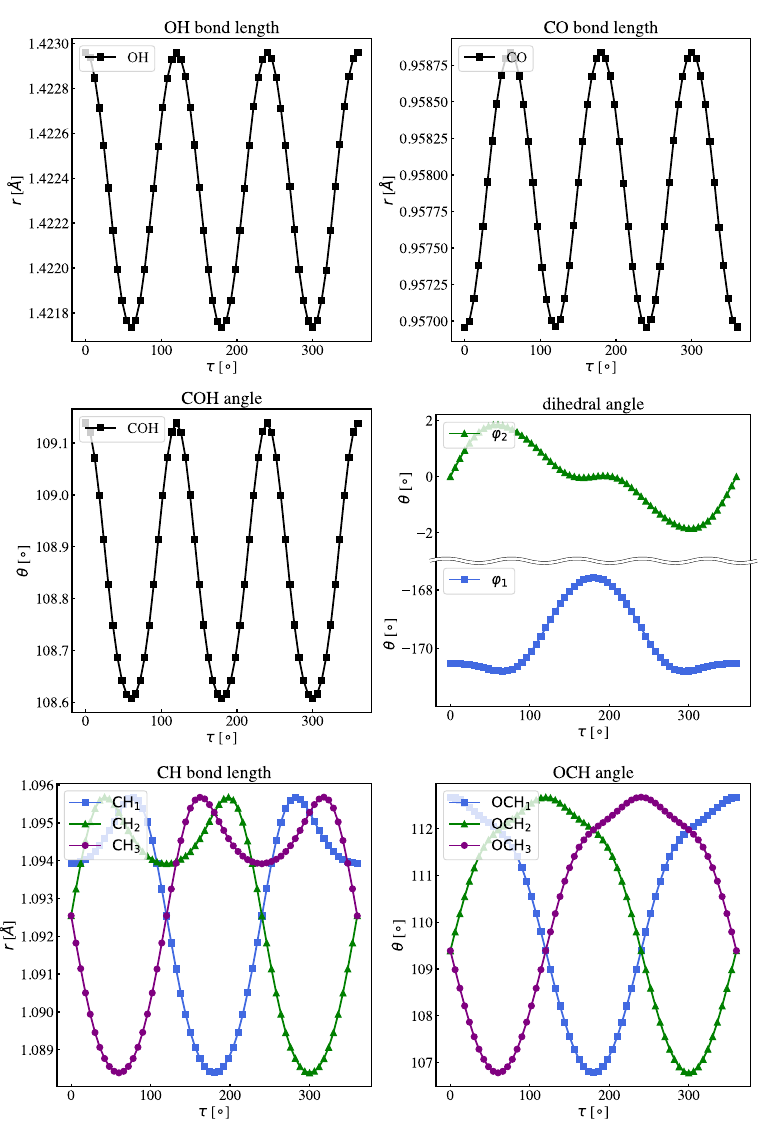}
\caption{%
\label{fig:xi_tau}
Internal coordinate values, $\xi^\mathrm{ref}_i(\tau)$, that minimize the potential energy along the $\tau$ torsional coordinate. 
}
\end{figure*}

We have imposed the exact fulfilment of these symmetry `features' on the computed c-normal coordinate coefficients as follows. 
For (i), we averaged the two C-H bonds and O-C-H angles that are, in principle, equivalent at the minima and saddle points. 
For (ii), the $\bos{\xi}^\tref(\tau_n)$ was relaxed in the interval $0^\circ \le \tau\le 60^\circ$, and the data set in the invterval $60^\circ < \tau\le 360^\circ$ was generated from the $0^\circ\le \tau\le 60^\circ$ values. 
Figure \ref{fig:xi_tau} shows the computed 
$\bos{\xi}^\tref(\tau_n)$ values and interpolated curves. 
The C--O and O--H bond lengths and the C--O--H angle feature 120$^\circ$ periodicity with three equivalent minima and maxima. The methyl C--H bonds and O--C--H angles have a $180^\circ$ periodicity (according to the minimum and saddle points). 
The necessary set of all of the coordinates appears in the region $0^\circ \le \tau\le 60^\circ$ (Figure \ref{fig:xi_tau}), and thus, we copied them to the region $60^\circ < \tau\le 360^\circ$ according to the expected symmetry relations.
To calculate the $\varphi_1$ and $\varphi_2$ angles, we first symmetrized
$\tau_1$, $\tau_2$, and $\tau_3$, and then, calculated $\varphi_1$ and $\varphi_2$ according to Eq.~\eqref{eq:lin_com}. 
A similar figure for \xitau\ in the region $0^\circ \le \tau\le 60^\circ$ has been reported in Fig.~2 of Ref.~\citenum{Sibert2005JCP}.

We define the following three types of reference structures for use in the later sections:
\begin{itemize}
  \item %
  ave-$\xi$ (averaged): 
    $\bos{\xi}^{\mathrm{ref}}
    =
    [\bos{\xi}^{\mathrm{ref}}(\tau_{\text{eq,1}})
    +
    \bos{\xi}^{\mathrm{ref}}(\tau_{\text{eq,2}})
    +
    \bos{\xi}^{\mathrm{ref}}(\tau_{\text{eq,3}})]/3$,
    where $i=1,2,3$ labels the three minima, 
    \emph{i.e.,} the average of the three equivalent minima is calculated; 
  \item %
    nonsym-$\xi(\tau)$ (no symmetrization): 
      $\bos{\xi}^\tref(\tau)$ obtained from direct computation but without imposing strict fulfilment of the exact symmetry relations (\emph{i.e.,} no elimination of numerical noise in double precision arithmetic);
  \item 
    sym-$\xi(\tau)$ (symmetrization (i) and (ii)): 
      $\bos{\xi}^\tref(\tau)$ obtained from direct computation and additional symmetrization to eliminate the effect of numerical noise due to the finite (double) precision arithmetic of the PES.
\end{itemize}

%
%

%
%
\vspace{0.5cm}
\paragraph{$\bos{L}(\tau)$: phase adjustment and symmetry considerations}
The GF method defines $\bos{L}$ only up to an arbitrary $(+1)/(-1)$ phase factor, since the $\bos{L}_k$ coefficient vectors are obtained as eigenvectors of the (non-symmetric) $\bos{G}^\stext\bos{F}^\stext(\tau_n)$ matrix, Eq.~\eqref{eq:GFL} (and normalization, Eq.~\eqref{eq:normalization}).
For interpolating the $\bos{L}(\tau_n)$ coefficients along the $\tau_n$ points, it is necessary to align the phase of the eigenvectors along the series of $\tau_n$ values.

To do this, we calculate the overlap of the eigenvectors at neighbouring $\tau_n$ points, and choose the phase of the eigenvector at the `next' point so that this overlap is closer to 1 than to $-1$, $\bos{L}^\text{T}_k(\tau_{n-1}) \bos{L}_k(\tau_{n})\approx 1 \ (k=1,\ldots,D_\stext)$. 
(For simplicity, we used here the transpose of the eigenvectors, $\bos{L}^\text{T}_k$, which was sufficient for the phase adjustment, and we only note that eigenvectors of a non-symmetric matrix are biorthogonal).
If the $\tau_n$ grid is sufficiently dense, then the phase adjustment can be unambiguously carried out. Difficulties arise when two eigenvectors ($\bos{L}_{k},\bos{L}_{k+1}$) are quasi-degenerate, and the order of the two vectors changes due to numerical error along the $\tau_n$ series of points (the eigenvectors are ordered according to decreasing order of $\lambda_k$). In this case, the $\bos{L}^\text{T}_{k}(\tau_{n-1}) \bos{L}_{k}(\tau_{n})$ overlap is too small, and then, the $k^{\mathrm{th}}$ and $(k+1)^{\mathrm{th}}$ eigenvectors are exchanged. 

Even if the symmetry is imposed at the $\bos{\xi}^\tref(\tau_n)$ reference structure, the symmetry is slightly broken in the $\bos{L}(\tau_n)$ eigenvectors due to the finite numerical precision. Scrutinizing symmetry is necessary, otherwise, the degeneracy of vibrational $E$ states is not well represented in finite-basis computations (Sec. \ref{subsec:ctau}). 

\begin{figure*}[!htbp]
\includegraphics[width=1.0\textwidth]{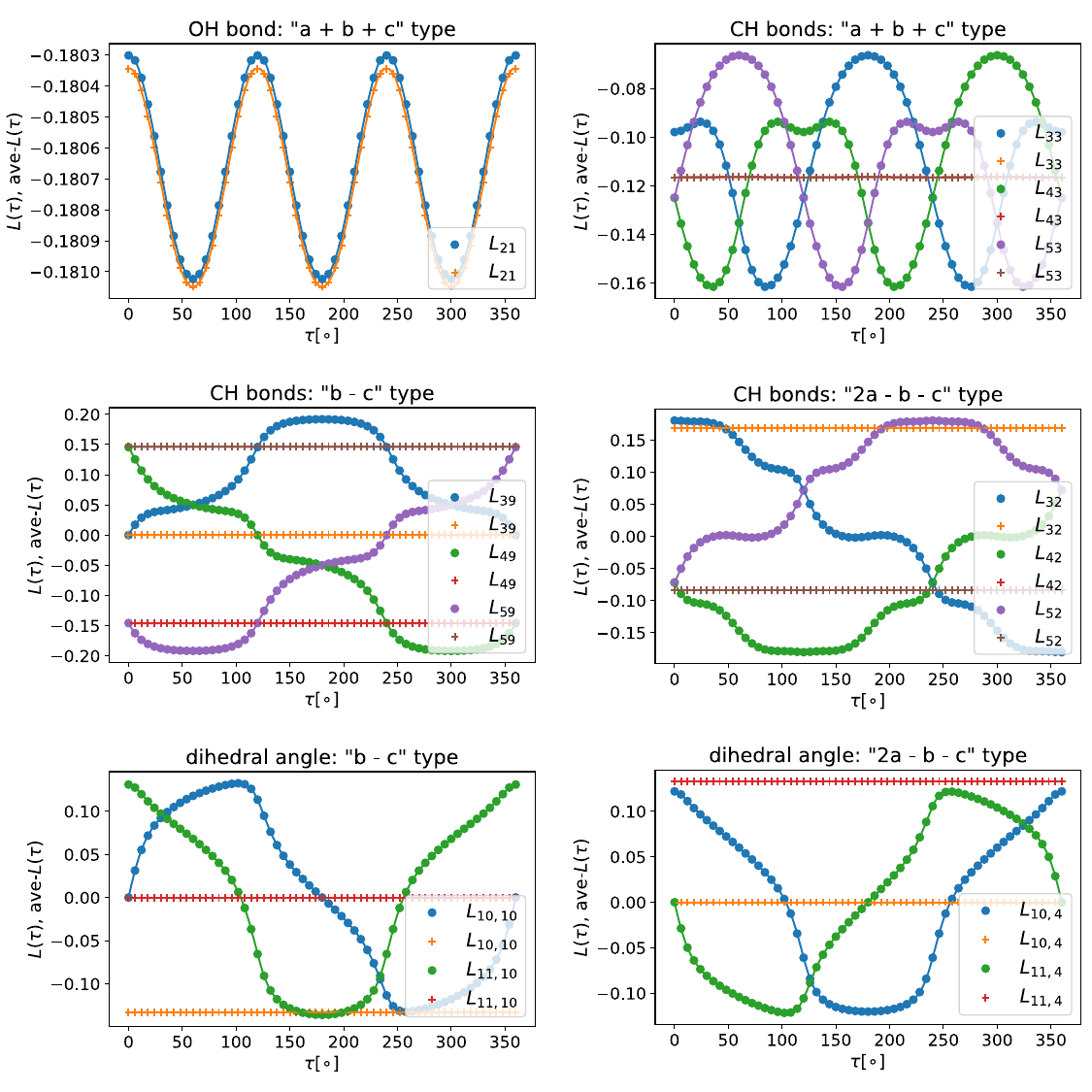}
\caption{\label{fig:L_tau} 
Torsional $\tau$ dependence of the $L_{ik}$ (dimensionless) curvilinear normal coordinate coefficients.
The sym-\Ltau\ and ave-\Ltau\ options are labelled with `$\bullet$' and `+', respectively. 
The index $i$ of $L_{ik}$ refers to the SAM internal coordinate ($\xi_i$), Eq.~\ref{eq:rhoxitau},
and $k$ labels the normal mode (Table \ref{tbl:harm_freq_aveL}).
}
\end{figure*}

We impose exact fulfilment of the symmetry relations on the $\bos{L}(\tau_n)$ coefficients as follows.
\begin{itemize}
  \item %
    sym-\Ltau\ (Figure~\ref{fig:L_tau}): similar to the sym-\xitau, the $C_\text{s}$ symmetry and $\tau$-periodicity is imposed based on data from the region $0^\circ\le \tau \le 60^\circ$. 
    The procedure is similar to that of sym-\xitau, with the additional phase 
    alignment step. 
    The associated harmonic frequencies are listed in Table S3 of the Supporting Information. The $a'$ ($a''$) label in this table refers to the irreducible representation (irrep) of the \Cs\ point group \cite{Bunker2006_molsym}, and corresponds here to symmetric (antisymmetric) vibrational modes. We note that we use small letters for the irreps of single-coordinate quantities following the convention of electronic structure theory using small letters for orbital symmetries (one-electron quantity) \cite{dyall2007introduction,reiher2014relativistic}.
  \item %
    ave-\Ltau\ (Figure~\ref{fig:L_tau}): To build sym-\Ltau, the $\bos{G}^\stext(\tau_n)$ and $\bos{F}^\stext(\tau_n)$ matrices are computed at each $\tau_n$. To construct ave-\Ltau, we first computed $\bos{G}^\stext(\tau_n)$ and $\bos{F}^\stext(\tau_n)$ with $\tau_n= 60^\circ\pm \delta, 180^\circ\pm \delta, 300^\circ\pm \delta$, where $ 0^\circ \le \delta \le 60^\circ$, 
    and then, calculated the average $\bar{\bos{G}}^\stext$ and $\bar{\bos{F}}^\stext(\tau_n)$ matrices averaged over these $\tau_n$ points.
    The $\bar{\bos{G}}^\stext$ and $\bar{\bos{F}}^\stext(\tau_n)$ matrices were used in the GF procedure. Then, the $\bos{L}_k$ eigenvectors correspond to irreps of $C_{3\text{v}}$(M). 
Linear combinations (carried by the $L_{ik}$ coefficients) of the internal coordinates are assigned to $a_1$ or $e$, as shown in Table \ref{tbl:harm_freq_aveL}.

Although the numerical GF procedure always has some numerical noise, we carried out the symmetrization to faithfully represent the $C_{3\text{v}}(\text{M})$ symmetry:
\emph{e.g.,} regarding the C--H bond distances, $(r_{\mathrm{CH}_1},r_{\mathrm{CH}_2},r_{\mathrm{CH}_3})$, the ratio of the linear combination coefficients of the dimensionless coordinates $q_3,q_9$ and $q_2$ (Table~\ref{tbl:harm_freq_aveL}) are exactly $(+1,+1,+1)$, $(0,+1,-1)$, and $(+2,-1,-1)$, respectively, within double precision arithmetic.
The harmonic frequencies obtained corresponding to the ave-\Ltau\ construction are listed in Table \ref{tbl:harm_freq_aveL}.
    \item sym-$L'(\tau)$, ave-$L'(\tau)$: When the numerical differentiation with respect to $\tau$ is carried out to obtain the kinetic energy coefficients (subsection~\ref{sec:clqrovib}) at $\tau_n$ \cite{Matyus2009JCP,Matyus2023CC},
    where $\tau_n$ is a grid point of the integration grid (subsection~\ref{sec:grid}), the $\tau$ dependence of sym-\Ltau\ (ave-\Ltau) is neglected and this approximation (used during the construction of the numerical kinetic energy operator coefficients) is shortly referred to as sym-$L'(\tau)$ (ave-$L'(\tau)$). This approximation was used for test purposes.
\end{itemize}

%
%

%
%
\begin{table}[!htbp]
\caption{%
Harmonic frequencies (\cm) computed with the ave-\Ltau\ GF procedure at three equivalent minima and PES13~\cite{Qu2013MP_CH3+OH}. 
The corresponding state label and description, the dominant internal coordinates, and symmetry label in the $C_{3\text{v}}(\text{M})$ MS group are also provided.
The assignment according to the \Ctv\ symmetry is explained in Sec.~S1 of the Supporting Information.
The $\nu_k\, (k=1,2,\cdots,11)$ labelling follows the literature convention, corresponding to \Cs\ symmetry, to have comparable assignments for the vibrational states (Sec.~\ref{subsec:compare_exp})
with earlier work. The $\nu_k \:(k=1,2,\cdots,8)$ vibrational modes are symmetric, whereas $\nu_k \:(k=9,10,11)$ are anti-symmetric for reflection to the COH plane. 
}
\label{tbl:harm_freq_aveL}
\begin{tabular}{@{}clccc@{}}
\hline\hline
state & description & coord & sym & $\tilde{\nu}$ \tabularnewline   
 \hline
$\nu_1$& $\nu$(\ce{OH}) & $r_\mathrm{OH}$ & $a_1$ & 3869.820  \tabularnewline 
$\nu_2$ & $\nu$(\ce{CH3})$_{\mathrm{asym}}$  & $r_\mathrm{CH}$ &$e$ & 3104.820 \tabularnewline 
$\nu_9$  & $\nu$(\ce{CH3})$_{\mathrm{asym}}$ & $r_\mathrm{CH}$&$e$ & 3104.820  \tabularnewline 
$\nu_3$& $\nu$(\ce{CH3})$_{\mathrm{sym}}$& $r_\mathrm{CH}$ &  $a_1$ & 3026.114  \tabularnewline 
$\nu_4$&$\delta$(\ce{CH3})$_{\mathrm{asym}}$ & $\varphi_2$ &  $e$ & 1510.890 \tabularnewline  
$\nu_{10}$ & $\delta$(\ce{CH3})$_{\mathrm{asym}}$ & $\varphi_1$ &$e$ & 1510.887  \tabularnewline 
$\nu_5$&$\delta$(\ce{CH3})$_{\mathrm{sym}}$ & $\theta_\mathrm{HCO}$ &  $a_1$ & 1475.296 \tabularnewline  
$\nu_6$&$\delta$(\ce{COH}) & $\theta_\mathrm{COH}$ &  $a_1$ & 1259.733 \tabularnewline  
$\nu_{11}$& $\rho$(\ce{CH3})$_{\mathrm{}}$& $\theta_\mathrm{HCO}$ &  $e$ & 1196.977  \tabularnewline
$\nu_7$  &$\rho$(\ce{CH3})$_{\mathrm{}}$ &$\theta_\mathrm{HCO}$ & $e$ & 1196.976  \tabularnewline
$\nu_8$& $\nu$(\ce{CO})$_{\mathrm{}}$&  $r_\mathrm{CO}$ & $a_1$ & 1066.481 \tabularnewline  
\hline\hline
\end{tabular}
\end{table}

Several components of sym-\Ltau\ and ave-\Ltau\ for the dominant internal coordinates are shown in Figure~\ref{fig:L_tau}.
For simplicity, only symmetric-unique components are provided, the $\bos{q}$ coordinates to which dominantly the O-C-H$_i \;(i=1,2,3)$ angles contribute behave similarly to C-H$_i \;(i=1,2,3)$ (shown in the figure). The coefficient vector of ave-\Ltau\ for the CH bonds shown in Figure~\ref{fig:L_tau} can be written as
\be\label{eq:matrix_q2_q3_q9}
\left(\begin{array}{c}
\Delta r_{\mathrm{CH_{1}}}\\
\Delta r_{\mathrm{CH_{2}}}\\
\Delta r_{\mathrm{CH_{3}}}
\end{array}\right)=\left(\begin{array}{ccc}
2l & 0 & l''\\
-l & l' & l''\\
-l & -l' & l''
\end{array}\right)\left(\begin{array}{c}
q_{2}\\
q_{9}\\
q_{3}
\end{array}\right),
\ee
where the first, second, and third columns of the $3 \times 3$ matrix of the right-hand side corresponds to the ($2a - b - c$), ($b - c$), and ($a + b + c$) types in Figure \ref{fig:L_tau}, respectively. 
We can see that the three columns are orthogonal but not orthogonal to other coefficient vectors of ave-\Ltau. 
If we assume the three vectors are normalized for simplicity,
\be\label{eq:xi_L_q}
\left(\begin{array}{c}
\Delta r_{\mathrm{CH_{1}}}\\
\Delta r_{\mathrm{CH_{2}}}\\
\Delta r_{\mathrm{CH_{3}}}
\end{array}\right)=\left(\begin{array}{ccc}
2/\sqrt{6} & 0 & 1/\sqrt{3}\\
-1/\sqrt{6} & 1/\sqrt{2} & 1/\sqrt{3}\\
-1/\sqrt{6} & -1/\sqrt{2} & 1/\sqrt{3}
\end{array}\right)\left(\begin{array}{c}
q_{2}\\
q_{9}\\
q_{3}
\end{array}\right),
\ee
where $q_i$ is the dimensionless c-normal coordinate  defined in Eq. \ref{eq:dxitau}, which is associated with $\nu_i$ listed in Table~\ref{tbl:harm_freq_aveL}.
The equations for $q_i$ are 
\bea\label{eq:q_xi}
q_{2} &=& \frac{1}{\sqrt{6}}\left(2\Delta r_{\mathrm{CH_{1}}}-\Delta r_{\mathrm{CH_{2}}}-\Delta r_{\mathrm{CH_{3}}}\right), \\ \nonumber
q_{9} &=& \frac{1}{\sqrt{2}}\left(\Delta r_{\mathrm{CH_{2}}}-\Delta r_{\mathrm{CH_{3}}}\right), \\ \nonumber
q_{3} &=& \frac{1}{\sqrt{3}}\left(\Delta r_{\mathrm{CH_{1}}}+\Delta r_{\mathrm{CH_{2}}}+\Delta r_{\mathrm{CH_{3}}}\right)\; .
\eea

Regarding comparison to previous work, the \Cs\ (point-group) based assignment is found in Table~2.1 of Ref.~\citenum{Moruzzi1995_ch3oh} and Table~V of Ref.~\citenum{Fehrensen2003JCP}. In Table~1 of Ref.~\citenum{Bowman2007JPCA}, $a'$ and $a''$ are replaced with $a_1$ and $a_2$ of \Ctv. 
In Ref.~\citenum{Sibert2005JCP}, \Ctv\ is retained in the Hamiltonian, and in fact, the degeneracy between several normal modes is found in their Table~V. The labels of the degenerate modes are the same as ours (Table~\ref{tbl:harm_freq_aveL}). Meanwhile, they do not explicitly show their normal coordinates' assignment in Table~V, and assign their eigenvalues using only $a_1$ and $a_2$ labels, even when the degenerate modes contribute to the state. 

%
%
Finally, we note that the core idea of a $\tau$-dependent reference configuration for the large-amplitude bending motion in a triatomic molecule was originally put forward by Hougen, Bunker, and Johns (HBJ)~\cite{Hougen1970JMS_rho_depend,Beardsworth1986JMS}. 
The HBJ Hamitonian was used and a computer implementation of the $\tau$-dependent GF method was reported in Refs.~\citenum{Viglaska2020JCP,Rey2023JCP} with ideas from Lie algebra regarding the symmetry of the reference configuration \cite{Szalay2018JCP_LAMPES}.
In the present work, we use not only a $\tau$-dependent reference configuration, but we choose this configuration to follow the minimal-energy path of the PES as it was done in Refs.~\citenum{Sibert2005JCP,Lauvergnat2014SA_methanol,Nauts2018MP_methanol}. Since in our computations, the PES is a general subroutine, we do not have the kinetic-energy operator in a simple analytic form (unlike the HBJ Hamiltonian), but we compute the KEO coefficients over a grid (according to Sec.~\ref{sec:clqrovib}). It can be expected that by considering the minimum energy path on the PES, the small-amplitude vibrations can be described more efficiently (using smaller basis sets). 

We also add that the minimum-energy path accounts only for the potential energy minimum along the path, but the nuclear kinetic energy is missing. We think that the most optimal $\tau$-following coordinate system may be defined along the instanton,\cite{Richardson2011JCP_instanton,MaWaAl16}
which is the minimum action path accounting for both the kinetic and potential energy of the nuclei, and we plan to test it in future work.

%
%
\section{Variational solution of the vibrational Schrödinger equation}\label{sec:metho}
We aim for the (quasi-)variational solution of the vibrational Schrödinger equation including the vibrational Hamiltonian, Eq.~\eqref{eq:Hfrearr}, 
\begin{align}
  \hat{H}^\text{v} \Psi = E \Psi \; ,
\end{align}
and to compute the ground and all vibrationally excited states up to an energy threshold. This task becomes computationally feasible by defining truncated direct product basis sets and corresponding pruned product integration grids. Efficient basis and grid truncation are possible in conjunction with curvilinear normal coordinates, which minimize the couplings of small-amplitude vibrations.

\subsection{Truncation of the direct-product basis}\label{subsec:plune_basis}
The vibrational wavefunction is straightforwardly expanded as a linear combination of the direct product of one-dimensional basis functions, 
\bea
&&\Psi^{\mathrm{DP}}_i\left(q_1, \ldots, q_{D_\text{s}}; \tau\right) \\ \nonumber
&=&\sum_{n_{1}=0}^{b_1} \ldots \sum_{n_{{D_\text{s}}}=0}^{b_{D_\text{s}}} \sum_{n_\tau=0}^{n_\tau} C_{n_{1}, \ldots, n_{{D_\text{s}}}, n_\tau}^i \prod_{j=1}^{D_\text{s}} \psi_{n_{j}}^{(q)} \left(q_j\right) \psi_{n_\tau}^{(\tau)}(\tau),
\eea
where the one-dimensional basis functions are most often constructed from orthogonal polynomials and $b_j$ is the largest degree of the polynomial in $\psi_{n_{j}}^{(q)}$. For the small-amplitude vibrations, we use harmonic oscillator basis functions, whereas for the $\tau$ torsional degree of freedom, Fourier basis function $\psi_{n_\tau}^{(\tau)}$ are used,
\be
1, \cos (\tau), \sin (\tau), \ldots, \cos \left(n_\tau \tau\right), \sin \left(n_\tau \tau\right), \quad \tau \in[0,2 \pi).
\ee
The direct-product representation increases the size of the basis set too rapidly.
A direct-product basis accounts for the coupling between all SAMs, but actually, the coupling between the SAMs is relatively small (by construction, Sec.~\ref{subsec:G_and_F}). We aim to prioritize a higher-order degree of the basis functions, instead of a high-order representation of the SAM couplings (that are small). We have implemented this idea by using a pruned product vibrational basis representation for the SAMs
\begin{align}
&\Psi^{\mathrm{NON-DP}}_i\left(q_1, \ldots, q_{D_\text{s}}; \tau\right) \nonumber\\ 
&= 
\sum_{f(n_1,\ldots,n_{D_\text{s}})\le b} \sum_{n_\tau=0}^{n_\tau} 
  C_{n_{1}, \ldots, n_{{D_\text{s}}}, n_\tau}^i 
  \prod_{j=1}^{D_\text{s}} \psi_{n_{j}}^{(q)}\left(q_j\right) \psi_{n_\tau}^{(\tau)}(\tau)
\label{eq:nondp}
\end{align}
with the so-called `pruning function' $f(n_1,\ldots,n_{D_\text{s}})$,
\be
f(n_1,\ldots,n_{D_\text{s}}) = \sum^{D_\text{s}}_i g_i(n_i) \geq0 \; .
\ee
Here $g_i(n_i)$ is a function to tune the weight of the number of basis functions included in the basis set for the coordinate $i$.
In the present study, the simplest possible function (identity) is used, 
\be\label{eq:alpha=1}
g_i(n_i) = n_i.
\ee
By choosing more complicated $g_i$ functions, we could give more `weight' to certain degrees of freedom in the basis representation, \emph{e.g.,} to include more functions for the low-energy SAMs.

The effect of fine-tuning $g_i(n_i)$ can improve the convergence of the energies and wave functions with respect to the total basis set size. 
In previous work on HCOOH~\cite{Avila2023PCCP_HCOOH}, the lowest harmonic wave number among the SAMs (632~\cm) was smaller than the wave number of the torsional fundamental vibration (673~\cm), so in that case, it was essential to increase the basis set size for this low-energy SAM \cite{Avila2023PCCP_HCOOH}. 
In methanol, the lowest-energy SAMs vibration wave number (1066~\cm, Table~\ref{tbl:harm_freq_aveL}) is two orders of magnitude larger than that of the LAM (9.1~\cm).
Furthermore, the basis sets with $n_i=1$ dominantly contribute to the lowest 100 states computed in the present work (Sec.~\ref{subsec:compare_exp}). Hence, the simplest pruning scheme, Eq.~\eqref{eq:alpha=1}, is expected to perform well in the present study. A more elaborate pruning scheme would be necessary to compute higher-excited states. It is an interesting idea to possibly combine the Smolyak scheme with basis sets contraction \cite{Wang2023JCP_H2O_dim,Kallullathil2023JCP_contraction,Felker2023_HF3,Simko2024JCP}. To the best of our knowledge, the combination of basis contraction and the Smolyak scheme was never reported, but it is ongoing work in our research group for a molecular complex.

%
%
\subsection{Truncation of the direct product integration grid \label{sec:grid}}
We employed the Smolyak quadrature scheme \cite{Smolyak1963SSSR,Petras2003_Numer_Math,avila2009JCP,avila2011JCP} for computing the matrix elements of the Hamiltonian with the basis functions defined in Section~\ref{subsec:plune_basis}.
This is an efficient scheme to obtain non-product grids as a linear combination of a product quadrature \cite{avila2009JCP,avila2011JCP}:
\bea
\hat{Q}(D, H) 
& =& \sum_{\boldsymbol{\sigma}(\bos{i}) \leq H} C_{i_1, i_2, \ldots, i_{D}} \hat{Q}_1^{i_1} \otimes \hat{Q}_2^{i_2}  \otimes \ldots \otimes \hat{Q}_{D}^{i_{D}},
\label{eq:smol}
\eea
where $D=3N-6$ is the number of vibrational degrees of freedom, in this work $D=12$ ($N=6$). Each term in the sum can be expressed as the (direct) product of the Smolyak quadrature for 11 SAMs and 1 trapezoidal quadrature for the LAM.
The analytic expression of $C_{i_1, i_2, \ldots, i_{D}}$ is shown in Refs.~\citenum{avila2009JCP,avila2011JCP}. The indices are defined as 
\bea\label{eq:index}
1 \leq i_{\chi}&\leq&  H-(D-1); \quad  1 \leq \chi \leq D_\text{s} \\  \nonumber
i_{\chi} &=&  M_{\tau}; \quad\quad\quad\quad\quad\quad\quad  \chi = D \\   \nonumber
\quad \chi&=&1(q_1),2(q_2),\ldots,D_\text{s}(q_{D_\text{s}}),D(\tau).
\eea
$\hat{Q}_\chi^{i_\chi}$ is an operator that generates a quadrature of a 1D polynomial $f(\rho_{\chi})$,
\be\label{eq:Q_ope}
\hat{Q}_\chi^{i_\chi} f\left(\rho_\chi\right)=\sum_{m=1}^{ M} w_{\chi, m} f\left(\rho_{\chi, m}\right) \approx \int^b_a w(x) f(x)\ \dd x \; .
\ee
In our study, the Gauss--Hermite quadrature \cite{Heiss2008JE_nested,avila2009JCP,Avila2013JCP_collocation,Avila2019JCP_methodology,sparse_grid} and trapezoidal quadratures are used for the SAMs and the LAM, respectively,
\be
\hat{Q}_\chi^{i_\chi}= 
\begin{cases}
\hat{Q}_{\mathrm{Her}}^{M=m_\chi\left(i_\chi\right)}, & \text{for } \chi=q_1, \ldots, q_{D_\text{s}}, \\ 
\hat{Q}_{\mathrm{trap}}^{M=M_\tau}, & \text{for } \chi= \tau. \
\end{cases}
\ee
The value of $m_\chi$ is decided so that all quadrature points of the quadrature rule $\hat{Q}_\chi^j$ can appear in the quadrature rule $\hat{Q}_\chi^{j+1}$. 
$d_\chi\left(i_\chi\right)$ is a maximum degree that is associated with $m_\chi\left(i_\chi\right)$.  
The quadrature in Eq.~\eqref{eq:Q_ope} can provide an exact value of the integration when the degree of the $f(\rho_{\chi})$ is $d_{\chi}(i_{\chi})$, and $d_{\chi}(i_{\chi})\geq 2 i_{\chi} -1$ is satisfied. The $m_\chi\left(i_\chi\right)$ and $d_\chi\left(i_\chi\right)$ values used in this work are listed as follows.
\par
%
\begin{center}
\begin{tabular}{@{}lcc ccc ccc ccc cc@{}}
\hline\hline 
  $i_\chi$                    & 1 & 2 & 3 & 4 & 5 & 6 & 7 & 8 & 9 &10 &11 &12 & $\ldots$ \\
  \hline 
  $m_\chi\left(i_\chi\right)$ & 1 & 3 & 3 & 7 & 9 & 9 & 9 & 9 &17 &19 &19 &19 & $\ldots$ \\
  $d_\chi\left(i_\chi\right)$ & 1 & 5 & 5 & 7 &15 &15 &15 &15 &17 &29 &29 &29 & $\ldots$ \\
\hline\hline\\
\end{tabular}
\end{center}

The advantage of the nested quadrature is that 
 a multivariate quadrature product grid $\hat{Q}^{j_1} \otimes \ldots \otimes \hat{Q}^{j_d}$ can share some of the grid points with other ones $\hat{Q}^{i_1}$ $\otimes \ldots \otimes \hat{Q}^{i_d}$, where $i_c \neq j_c$ for more than one coordinate. 
$H$ is the grid pruning parameter of the Smolyak quadrature, 
\bea
\bos{\sigma}(\bos{i})&=&\sum_{\chi}^D s^{\chi}(i_{\chi})\le H, \\ \nonumber
 s^\chi\left(i_\chi\right)&=&\left\{\begin{array}{l}   i_{\chi}, \text { for } \chi=q_1, \ldots, q_{D_s},   \\ 
 1, \text { for } \chi=\tau. \end{array}\right.
\eea
The parameter that provides the largest number of grid points for a SAM, $K$ (Eq. \ref{eq:index}), is defined by $K=H-(D-1)$. In this work, we found $H=b+14$ to perform well.

In the GENIUSH program, the kinetic energy coefficients are computed over the integration grid. At every grid point, we compute the rotational \tvec-vectors,  Eq.~\eqref{eq:rot_t}, 
and the vibrational \tvec-vectors, Eq.~\eqref{eq:vib_t}. 
The vibrational \tvec-vectors are computed by finite differences, using the central difference formula with the $\Delta \rho_k=1.0 \times 10^{-4}$ step size.
Then, the $\bos{g}$ matrix is assembled according to Eq. \eqref{eq:small_g}, from which $\bos{G}=\bos{g}^{-1}$ and $\detg=\det\bos{g}$ are computed. 
For the fully rearranged form of the Hamiltonian, higher-order (up to third-order) coordinate derivatives are also needed, and the $B_{l}$ and $U$ terms are evaluated by repeated use of the chain rule and the coordinate derivatives \cite{Matyus2009JCP}.
The higher-order coordinate derivatives are computed with quadruple precision arithmetic, which becomes inefficient for high-dimensional systems (with many grid points). 
Alternatively, the Figure~\ref{fig:fortran_code} approach or an automated analytic differentiation, by repeated use of the chain rule \cite{Yachmenev2015JCP_ana_der}, could be used for a more efficient computation of sufficiently precise coordinate derivatives. Furthermore, if only the vibrational block of the $\bos{G}$ matrix is required, the $\bos{s}$-vector formalism \cite{Sorensen1979_LAM} may be more efficient instead of using \tvec-vectors, since the $\bos{G}\in\mathbf{R}^{(3N-6)\times(3N-6)}$ vibrational block can be directly computed with $\bos{s}$-vectors without matrix inversion.

%
%
\subsection{Sigmoid functions and curvilinear coordinate ranges}
The dimensionless normal coordinates can take values in the $q\in(-\infty,\infty)$ interval, and correspondingly, the integration grid points are in this interval. If we used these values in Eq.~\eqref{eq:xi_Lq}, we may obtain $\bos{\xi}$ `primitive' internal coordinate values outside the coordinate range of bond distances, bond angles.
To ensure that the internal coordinate values are in their mathematically correct interval, \emph{i.e.,} bond distance, $r\in [0,\infty)$; bond angle, $\theta \in [0,\pi]$; torsional coordinate, $\tau \in [0,2\pi)$; we used the `mapping functions' defined in Ref.~\citenum{Daria2022JMS_HCOOH}. 
The parameters in the mapping functions are tuned so that small variations in the precise choice of the mapping function itself do not significantly affect the vibrational energies.
In our largest computations ($b=8$, Sec.~\ref{subsec:compare_exp}), the largest (smallest) grid points corresponding to the dimensionless normal coordinates are about $\pm 4.5$. From the conversion through the $L$ transformation matrix, the coordinate range in the internal coordinate, $\Delta\bos{\xi}$, is about $\pm 0.81$ in bohr and radian (for distances and angles) from the reference coordinate value, $\bos{\xi}^\text{ref}$. 
In this computation, the mapping functions are necessary for some degrees of freedom, and they will be essential for larger basis and grid sizes.

\section{Numerical results and discussion}\label{sec:results}
First of all, we note that the expected accuracy of the vibrational band origins is limited to a few cm$^{-1}$ (due to the limited accuracy of the electronic energies, the PES fitting, and the non-relativistic and BO approximations). Nevertheless, we show additional digits in several tables  due to methodological reasons. The manuscript and Supporting Information extensively discuss the tight convergence of degenerate pairs, which is relevant for planned rovibrational computations and direct comparison with high-resolution microwave and infrared spectra of rotational intervals.

%
%
\subsection{Effect of the reference structure}\label{subsec:ctau}
First, we have studied the numerical effect of the relaxation of the reference structure in the case of a 1D model. 
We have constructed the 
reduced dimensionality models with `reduction in the $\bos{g}$ matrix' \cite{Matyus2009JCP,Matyus2023CC}. In the 1D case, the different 1D vibrational models correspond to different approximations to the full-dimensional problem. 

Table~\ref{tbl:1D_sym} shows the 1D vibrational energies obtained with the ave-$\xi$ (averaged over the equilibrium) and sym-\ctau ($\tau$ following, symmetrized) reference structures. Furthermore, the effect of the symmetrization with respect to a non-symmetrized (nonsym) $\tau$-following model is also shown in the table. 
In the 1D model, the difference between sym-\ctau and nonsym-\ctau is very small, but it becomes significant in the full-dimensional 12D model. 

Table \ref{tbl:12D_b=4} summarizes 12D results obtained using various \xitau, \Ltau\ options (see also Sec.~\ref{subsubsec:LAM-following}) and small basis and grid sizes.
It can be observed in this table that the lack of symmetrization (symmetry correction for the numerical noise) increases non-negligible numerical errors in degenerate energy pairs (for a given basis size). On the other hand, if sym-\ctau is used, then degeneracies can be (numerically) converged better than 0.01 \cm\ with moderate basis sizes. We used this sym-\ctau option for large-scale computations reported in Sec.~\ref{subsec:benchmark}.

%
%
\begin{table}
\caption{%
  Vibrational energies of CH$_3$OH, $\tilde\nu$ in \cm, referenced to the zero-point vibrational energy (\#1) using the 1D torsional model defined with the ave-$\xi$ and sym-\ctau reference structures using GENIUSH and PES13 \cite{Qu2013MP_CH3+OH}. Comparison with nonsym-\ctau results are also shown, $\delta\tilde{\nu}_\text{sym}=\tilde{\nu}_{\text{sym}}-\tilde{\nu}_{\text{nonsym}}$. 
  25 Fourier-DVR grid points were used to solve the vibrational Schrödinger equation.
}
\label{tbl:1D_sym}
\begin{tabular}{@{}crrrr@{}}
\hline\hline
  \#
 & \multicolumn{1}{c}{$\tilde\nu_{\text{ave-}\xi}$}
 & \multicolumn{1}{c}{$\tilde\nu_{\text{sym-}\xi(\tau)}$}
 & \multicolumn{1}{c}{$\delta\tilde\nu_{\mathrm{sym}}$}
 & \multicolumn{1}{c}{$\tilde\nu_\text{expt}$ \cite{Moruzzi1995_ch3oh}} 
 \tabularnewline
 \hline
1 & 235.155 & 120.465 &$ -6\times 10^{-4} $& \tabularnewline
2 & 13.558 & 9.909 & $2\times 10^{-4}$ & 9.1\rule[0mm]{0mm}{5mm} \tabularnewline
3 & 13.558 & 9.909 & $-4\times 10^{-4 }$& \tabularnewline
4 & 168.864 & 199.462 &$ 5\times 10^{-4}$ & 208.9\rule[0mm]{0mm}{5mm} \tabularnewline
5 & 168.864 & 199.462 &$ 1\times 10^{-4}$ & \tabularnewline
6 & 271.206 & 288.262 & $-4\times 10^{-4}$ & 294.5\rule[0mm]{0mm}{5mm} \tabularnewline
7 & 305.663 & 342.399 &$ 7\times 10^{-5}$ & 353.2\rule[0mm]{0mm}{4mm}\tabularnewline
8 & 475.820 & 502.659 &$ 1\times 10^{-3}$ & 510.3\rule[0mm]{0mm}{5mm} \tabularnewline
9 & 475.820 & 502.659 &$ -3\times 10^{-3} $& \tabularnewline
10 & 719.407 & 745.858 &$ 2\times 10^{-3} $& 751.0\rule[0mm]{0mm}{5mm} \tabularnewline
11 & 719.407 & 745.858 &$ -5\times 10^{-3}$ & \tabularnewline
12 & 1018.897 & 1046.138 &$ -1\times 10^{-2} $& 1046.7\rule[0mm]{0mm}{5mm}\tabularnewline
13 & 1019.021 & 1046.447 &$ 9\times 10^{-3} $& \rule[0mm]{0mm}{4mm}\tabularnewline
14 & 1373.709 & 1402.634 &$ 6\times 10^{-3} $&\rule[0mm]{0mm}{5mm}  \tabularnewline
15 & 1373.709 & 1402.634 &$ -1\times 10^{-2} $& \tabularnewline
16 & 1783.393 & 1814.412 &$ 5\times 10^{-4}$ & \rule[0mm]{0mm}{5mm} \tabularnewline
17 & 1783.393 & 1814.412 &$ -1\times 10^{-2}$ & \tabularnewline
18 & 2247.897 & 2281.419 &$ -2\times 10^{-2}$ & \rule[0mm]{0mm}{5mm} \tabularnewline
19 & 2247.897 & 2281.444 &$ 4\times 10^{-3}$ & \tabularnewline
20 & 2767.167 & 2803.596 &$ 2\times 10^{-3}$ & \rule[0mm]{0mm}{5mm} \tabularnewline
\hline\hline
\end{tabular}
\end{table}

%
%
\begin{table*}
\caption{%
  Vibrational energies referenced to the ZPVE, in cm$^{-1}$, from a 12D computation using different coordinate choices (reference geometry and transformation matrix) and 
  the PES13~\cite{Qu2013MP_CH3+OH} 
  with one exception, where the PES07~\cite{Bowman2007JPCA} was used as indicated in the column heading. 
  The basis and integration grid parameters were $b=4,H=18$ for the SAMs, and $n_{\tau}=32,M_{\tau}=54$ for the LAM. 
  We note that the $L'$ cases correspond to approximate kinetic energy coefficients (Sec.~\ref{subsubsec:LAM-following}).
  The unacceptably large numerical splittings of degenerate pairs, more than 0.01 \cm, are indicated in bold.
}
 \label{tbl:12D_b=4}
\begin{tabular}{@{}crrrrrrrr@{}}
\hline\hline
&  ave-$\xi$ & nonsym-\xitau & sym-\xitau & sym-\xitau & sym-\xitau 
& sym-\xitau & sym-\xitau & \tabularnewline
&  ave-$L$ & ave-$L$ & ave-$L$ & sym-$L'(\tau)$ & ave-$L'(\tau)$ & ave-$L'(\tau)$ & ave-\Ltau & exp. \cite{Moruzzi1995_ch3oh} \tabularnewline
 & PES13 & PES13 & PES13 & PES13 & PES07 & PES13 & PES13 & \tabularnewline
 \hline
 1 & 11111.511 & 11109.940 & 11109.942 & 11110.805 & 11093.354 & 11109.924 & 11109.947& \tabularnewline
2 & 9.181 & 9.126 & 9.128 &\textbf{ 9.006} & 9.187 & 9.180 & 9.126& 9.1\rule[0mm]{0mm}{4mm}\tabularnewline
3 & 9.181 & 9.135 & 9.130 & \textbf{9.054} & 9.190 & 9.189 & 9.130 &  \tabularnewline

4& 207.484  & \textbf{207.972} & 207.969 & \textbf{207.295} & 205.597 & 208.021 & 207.976&208.9\rule[0mm]{0mm}{5mm} \tabularnewline
5& 207.485  & \textbf{207.985 }& 207.973 & \textbf{207.491} & 205.602 & 208.025 & 207.980&  \tabularnewline
6&  292.192 & 292.466 & 292.461 & 291.989 & 291.001 & 292.527 & 292.467&294.5\rule[0mm]{0mm}{5mm} \tabularnewline
7&  352.975 & 353.574 & 353.572 & 353.357 & 347.711 & 353.423 & 353.583& 353.0\rule[0mm]{0mm}{4mm} \tabularnewline
8 & 509.053 & \textbf{509.452} & 509.459 & \textbf{508.879} & \textbf{503.971} & 509.492 & 509.468& 510.3\rule[0mm]{0mm}{5mm} \tabularnewline
9 & 509.054 & \textbf{509.469} & 509.459 & \textbf{509.302} & \textbf{504.009} & 509.494 & 509.469&  \tabularnewline
10& 749.388  & 749.726 & 749.719 & \textbf{748.183} & \textbf{742.300} & \textbf{749.747} & 749.727& 751.0\rule[0mm]{0mm}{5mm} \tabularnewline
11 & 749.390 & 749.729 & 749.724 & \textbf{749.413} & \textbf{742.437} & \textbf{749.794} & 749.734&  \tabularnewline
12& 1044.472  & 1044.715 & 1044.713 & 1043.197 & 1034.979 & 1044.696 & 1044.723&1046.7\rule[0mm]{0mm}{5mm} \tabularnewline
13& 1044.953  & 1045.114 & 1045.114 & 1045.292 & 1035.246 & 1045.233 & 1045.126& \rule[0mm]{0mm}{4mm}\tabularnewline
14& 1053.464  & 1049.923 & 1049.926 & 1048.936 & 1044.024 & 1049.860 & 1049.924& \rule[0mm]{0mm}{4mm}\tabularnewline
15& \textbf{1062.609} & 1058.928 & 1058.927 & \textbf{1058.510} & 1051.977 & \textbf{1058.958} & 1058.929& \rule[0mm]{0mm}{5mm} \tabularnewline
16&  \textbf{1079.291} & 1058.933 & 1058.931 & \textbf{1058.550} & 1051.980 & \textbf{1058.983} & 1058.934& \tabularnewline
17&  1084.174 & 1071.686 & 1071.683 & 1062.619 & 1083.964 & 1071.719 & 1071.696& \rule[0mm]{0mm}{5mm} \tabularnewline
18& \textbf{1172.175} & 1076.312 & 1076.310 & \textbf{1071.985} & \textbf{1088.330} & \textbf{1076.344} & 1076.316& \rule[0mm]{0mm}{5mm} \tabularnewline
19&\textbf{1181.812} & 1076.319 & 1076.311 & \textbf{1072.397} & \textbf{1088.365} & \textbf{1076.359} & 1076.319& \tabularnewline
20& 1258.160 & 1166.602 & 1166.606 & 1173.188 & 1163.072 & 1166.668 & 1166.613& \rule[0mm]{0mm}{4mm}\tabularnewline
\hline\hline
\end{tabular}
\end{table*}

%
%
\subsection{Effect of the coefficient vectors}\label{subsec:res:Ltau}
First of all, the numerical precision of our GF implementation is examined by comparing the harmonic frequencies at the three equilibrium minima (Table S3 of the Supporting Information).
For the Hessian computation, the 8$^\text{th}$-order central field differentiation (Section~\ref{subsec:G_and_F}) was used with the step sizes of 0.1 radians for $\varphi_1,\varphi_2,\mathrm{and} \; \tau$, and 0.06 (bohr and radian) for the remaining coordinates.

Table~\ref{tbl:12D_b=4} highlights that it is important to ensure the numerical fulfilment of the $C_{3\text{v}}(\text{M})$ symmetry in the c-normal coordinate definition to high precision, \emph{i.e.,} at the level of the $\bos{\xi}^\tref$ reference structure and the \Ltau\ coefficient vector, otherwise degenerate energies converge slowly with respect to the basis size.
The ave-$L'(\tau)$ coefficients can reproduce the results of the ave-\Ltau\ well for several states, but the largest deviation is 0.16~\cm\ (7th state). This convergence level is not sufficiently good for benchmark computations, so we use ave-\Ltau\ in the later large-scale computations. 

By comparison of the vibrational energies computed with PES07 \cite{Bowman2007JPCA} and PES13 \cite{Qu2013MP_CH3+OH} (not yet perfectly converged with $b=4$ in Table~\ref{tbl:12D_b=4}) and experiment, we observe that PES13 provides closer values to the experimental values.
At the same time, no significant difference is found between ave-$L$ and ave-\Ltau, which can be explained by the relatively small torsional barrier (349.6~\cm\ \cite{Qu2013MP_CH3+OH}).
In fact, Figure~\ref{fig:L_tau} shows that the ave-$L_{ik}(\tau)$ coefficients corresponding to the
C--H bond distances and the dihedral angles have only a very weak $\tau$ dependence. The ave-$L_{ik}(\tau)$ coefficient corresponding to the O--H bond distance slightly depends on $\tau$. 
All $L_{ik}(\tau)$s fulfill $\tau=[0,2\pi]$ periodicity, \emph{i.e.,} $L_{ik}(0)=L_{ik}(2\pi)$, except for the `2a - b - c' and `b - c' cases of the CH bonds, for which the $L_{32},L_{42},L_{49},L_{52},L_{59}$ coefficients have opposite signs at $\tau=0$ and $2\pi$, \emph{e.g.,} $L_{32}(0)=-L_{32}(2\pi)$. These special cases are $\tau=[0,4\pi]$ periodic, which is reminiscent to the half-integer case of the angular momentum in double groups.
Actually, the necessity of using the double group to classify the torsional wave function ($\Phi_{\text {tor }}=[1 /(2 \pi)]^{1 / 2} \exp \left(i k \gamma\right)$) has been noted in Sec.~15.4.4 of Ref.~\citenum{Bunker2006_molsym}. The sign of the symmetry transformation property is changed when $k$ is an odd value and $\gamma$ is replaced by $\gamma+2\pi$. The torsional splitting pattern of methanol was investigated by Di Lauro and Lattanzi \cite{Lattanzi2005MP,Di_Lauro2020CJP,Di_Lauro2020_Rotational_structure}.
Possible further developments of the sym-$L(\tau)$ representation towards double groups is left for future work.

At first sight, the poorly reproduced degeneracies in the 12D computation using \Ltau\ (Table~\ref{tbl:12D_b=4}) were surprising (after checking our GF code carefully), and we think that it is due to the (lack of) symmetry in \Ltau.
From Figure~\ref{fig:L_tau}, ave-\Ltau\ satisfies the $C_{3\text{v}}$(M) for all $\tau$ for the C-H bond distances and the dihedral angles. Only symmetry-allowed $\xi_i$ coordinates contribute to $q_k$. 
Meanwhile, for sym-$L(\tau)$, $C_\text{s}$-type symmetry can be found at the three equilibrium minima and saddle points of the C--H bond distances and the dihedral angles, but the symmetry is $C_1$ for other $\tau$ values.
This feature may affect the convergence of degenerate pairs.

Table~\ref{tbl:harm_freq_aveL} shows the ave-\Ltau\ harmonic frequencies at three equivalent minima.
Since the $\tau$-dependence is relatively small, the harmonic frequencies at other $\tau$ values are not shown. The symmetry label is not changed in other $\tau$ values because \Ctv\ is fulfilled in all $\tau$\, and the labels are uniquely determined, as explained in Sec. S1 of the Supporting Information.
%

%
%

%
%
\subsection{Benchmark vibrational energies and assignment of the vibrational wave function}\label{subsec:benchmark}

The vibrational wave functions are assigned by identifying the dominant basis contributions corresponding to the curvilinear normal coordinate representation with the sym-$\xi(\tau)$ and ave-$L$ parameterization.
Furthermore, the wave functions are assigned to irreducible representations of the $C_{3\text{v}}$(M) molecular symmetry group \cite{Bunker2006_molsym} based on the torsional and SAM symmetry properties (Sec. S1 of the Supporting Information).
Comparisons with vibrational band origins derived from experiments and earlier computations are also listed in the tables. 
In a few cases, deviating assignments are indicated in the footnotes to the table.

%
%
\subsubsection{Assignment of the torsional states}\label{subsec:tor_quant}
In the case of the SAMs, the number of nodes corresponds to the vibrational quantum number of the motion, but the number of nodes and the quantum number are not simply associated in the case of the torsional motion which extends over several minima. The quantum number (label) of the purely torsional states, the dominant Fourier basis function, and the number of nodes are listed in Table \ref{tbl:tor_quant}. The 1D torsional wavefunction is shown in Fig.~S3 of the Supporting information.

\begin{table}[ht]
\caption{%
Characterisation of the torsional excitation. Number of nodes and the dominant basis functions obtained from the computation shown in Table \ref{tbl:b=8_assign1} for the torsional quantum number, $\nu_{\tau}$, up to $\nu_{\tau}=2$.
}\label{tbl:tor_quant}
\begin{tabular}{@{}ccccc@{}}
\hline\hline
 & \multicolumn{2}{c}{\# of nodes} & \multicolumn{2}{c}{Dominant basis functions}\tabularnewline
$\nu_{\tau}$ & $a$ & $e$ & $a$ & $e$\tabularnewline
\hline
0 & 0 & 2 & 1 & \{sin($\tau$), +sin($2\tau$)\}, \{cos($\tau$), --cos($2\tau$)\}\tabularnewline
1 & 6 & 4 & sin($3\tau$) & \{sin($2\tau$), --sin($\tau$)\}, \{cos($2\tau$), +cos($\tau$)\}\tabularnewline
2 & 6 & 8 & cos($3\tau$) & \{sin($4\tau$), +cos($4\tau$)\}, \{sin($4\tau$), --cos($4\tau$)\}\tabularnewline
\hline\hline
\end{tabular}
\end{table}

A feasible approach to assign torsional quantum numbers (unambiguous torsional labels) involves initially allocating a torsional quantum number to the pure torsional states. Subsequently, the principal Fourier basis functions that contribute to each pure torsional state are identified. These Fourier basis functions can then be used to assign torsional quantum numbers to the SAMs-LAM coupling states. 
For instance, the dominant basis functions and the coefficients in the 53rd state (Supporting~Information) are as follows:
\be
-0.51\sin(2n\tau)\psi_{n_6=1}(q_6) + 0.41\cos(3n\tau)\psi_{n_11=1}(q_{11}) + 0.34\sin(3n\tau)\psi_{n_7=1}(q_{7}) + 0.26\sin(n\tau)\psi_{n_6=1}(q_6)
\ee
where only the basis functions with a non-zero excitation number are explicitly written. The second term with $\cos(3\tau)$ is assigned to $v_{\tau}=2$ and the other terms with $\sin(3\tau)$ and \{$\sin(2\tau),-\sin(\tau)$\} are assinged to $v_{\tau}=1$ (Table \ref{tbl:tor_quant}). In our energy region, the largest torsional quantum number $v_{\tau}=5$ is found for the 72nd-73rd states, in which $\sin(8\tau)$ and $\cos(8\tau)$ give dominant contributions.  

\begin{table*}[!htbp] 
\begin{center}     
 \caption{
 \small{
 Vibrational states of CH$_3$OH in 12D computed with the GENIUSH-Smolyak program and the PES13 potential energy surface \cite{Qu2013MP_CH3+OH}. 
 Vibrational energies, $\tilde{\nu}$ in \cm, referenced to the zero-point vibrational energy (\#1) and assignments obtained using the $b=8$ basis set and the sym-\xitau\ and ave-$L$ curvilinear set of normal coordinates. Comparison with earlier computational results and experimental vibrational band origins are also shown. 
 SAMs: assignment of the curvilinear normal modes $1_n, 2_n, ..., 11_n$ $(n=0,1,\ldots)$ in Table~\ref{tbl:harm_freq_aveL},
 zero excitation ($n=0$) is not shown.  `$[\left.\ldots\right]$' labels the largest contribution(s) from strongly mixed states. 
 Sym: irrep label of $C_{3\text{v}}(\text{M})$ according to Table~\ref{tbl:harm_freq_aveL} and corresponding discussion.
 All degeneracies for the $E$ states converged better than 0.01~\cm, except for the three pairs listed in the footnotes. 
 $\delta=\tilde{\nu}_{b=7}-\tilde{\nu}_{b=8}$, in \cm, is shown for assessment of the convergence.
Refs.~\citenum{Fehrensen2003JCP,Bowman2007JPCA,Sibert2005JCP} used the reaction-path Hamiltonian in variational,  MULITMODE (PES07) and perturbation theory computations, respectively. The variational Smolyak computations\cite{Lauvergnat2014SA_methanol,Nauts2018MP_methanol} also used PES07.
 }
}
\label{tbl:b=8_assign1}
\scalebox{0.9}{
 \begin{tabular}{@{}rrrrr r rr rrrc@{}}
 \hline\hline
 \# & 
 $\nu_{\tau}$ & 
 SAMs & 
 irrep & 
 $\tilde{\nu}$ &  $\delta$ &  Ref.~\citenum{Fehrensen2003JCP} & Ref.~\citenum{Sibert2005JCP} & Ref.~\citenum{Bowman2007JPCA} & Ref.~\citenum{Lauvergnat2014SA_methanol,Nauts2018MP_methanol}  & 
 
 \multicolumn{2}{c}{$\tilde{\nu}_\text{Expt.}$}
 \tabularnewline
 \hline
 1 & 0 &  & $A_1$ & 11108.023 & (0.02) &                                    &  &  &  &                                                                              \tabularnewline
 2-3 & 0 &  & $E$ & 9.118 & (0.0) &                                     8.2 &  & 8.7 & 9.15 &                                                                  9.122 & \citenum{Moruzzi1995_ch3oh}\tabularnewline
 4-5 & 1 &  & $E$ & 208.0 & (0.0) &                                        216.7 &  & 205.3 & 205.3 &                                                          208.9 & \citenum{Moruzzi1995_ch3oh}\tabularnewline
 6 & 1 &  & $A_2$ & 292.4 & (0.0) &                                        297.8 &  & 267.0 & 290.7 &                                                          294.5 & \citenum{Moruzzi1995_ch3oh}\tabularnewline
 7 & 2 &  & $A_1$ & 353.6 & (0.0) &                                        363.5 &  & 388.2 & 347.5 &                                                          353.2 & \citenum{Moruzzi1995_ch3oh}\tabularnewline
 8-9 & 2 &  & $E$ & 509.3 & (0.0) &                                        517.2 &  & 509.3 & 503.5 &                                                          510.3 & \citenum{Moruzzi1995_ch3oh}\tabularnewline
 10-11 & 3 &  & $E$ & 749.5 & (0.0) &                                      758.7 &  & 762.3 & 741.8 &                                                            751.0 & \citenum{Moruzzi1995_ch3oh}\tabularnewline
 12 & 0 & 8$_1$ & $A_1$ & 1036.3 & (0.3) &                                 1064.8 & 1031.0 & 1026.3 &  &  1034.4                                                            & \citenum{Lees2002PRA}  \tabularnewline
 13 & 3 &  & $A_2$ & 1044.2 & (0.0) &                                      1057.7 &  & 1017.8 & 1034.5 &                                                               1046.7 & \citenum{Moruzzi1995_ch3oh}\tabularnewline
 14 & 4 &  & $A_1$ & 1045.0 & (0.0) &                                      1058.1 &  & 1078.0 &  &                                                                 1047.6 & \citenum{Moruzzi1995_ch3oh}\tabularnewline
 15-16 & 0 & 8$_1$ & $E$ & 1045.6 & (0.3) &                                1073.2 & 1039.4 & 1036.8 &  &                                                                     1042.6 & \citenum{Lees2002PRA}\tabularnewline
 17 & 0 & 6$_1$,7$_1$,11$_1$ & $A_1$ & 1059.4 & (0.3) &                    1129.1 & 1074.0 & 1079.9 &  &                                                                           1074.7  & \citenum{Lees2002PRA}\tabularnewline
 18-19 & 0 & 6$_1$,7$_1$,11$_1$ & $E$ & 1064.0 & (0.3) &                   1120.7 & 1078.6 & 1087.5 &  &                                                                                  1079.3 & \citenum{Lees2002PRA}\tabularnewline
 20-21 & 0 & 11$_1$, 10$_1$ & $E$ & 1153.8 & (0.3) &                       1234.2 & 1152.4 & 1151.1 &  &                                                                        1156.5 & \citenum{Lees2002PRA}\tabularnewline
 22 & 0 & 11$_1$, 10$_1$ & $A_2$ & 1163.5 & (0.3) &                        1242.4 & 1159.9 & 1156.5 &  &                                                                             1164.0 & \citenum{Lees2002PRA}\tabularnewline
 23-24 & 1 & 8$_1$ & $E$ & 1242.2 & (0.3) &                                1196.9 &  & 1178.2 &  &                                                                            & \tabularnewline
 25-26 & 1 & 6$_1$,7$_1$,11$_1$ & $E$ & 1285.5 & (0.3) &                    &  &  &  &                                                                                               & \tabularnewline
 27 & 1 & 6$_1$,7$_1$,11$_1$ & $A_1$ & 1309.5 & (0.3) &                     &  &  &  &                                                                                              & \tabularnewline
 28 & 1 & 6$_1$,7$_1$,11$_1$ & $A_2$ & 1322.5 & (0.3) &                    1408.8 & 1321.0 & 1321.9 &  &                                                                           1320.6 & \citenum{Lees2004JMS}\tabularnewline
 29 & 1 & 8$_1$ & $A_2$ & 1328.6 & (0.3) &                                 1279.5 &  & 1234.9 &  &                                                                           & \tabularnewline
 30-31 & 0 & 6$_1$ & $E$ & 1336.4 & (0.2) &                                1417.5 & 1345.9 & 1341.9 &  &                                                                     1335.2 & \citenum{Lees2004JMS}\tabularnewline
 32 & 1,2 & 6$_1$,7$_1$,11$_1$ & $A_1$ & 1372.6 & (0.3) &                     &  &  &  &                                                                                             1369.7 & \citenum{Lees2004JMS}\tabularnewline
 33 & 2 & 8$_1$ & $A_1$ & 1387.2 & (0.3) &                                  &  &  &  &                                                                                1388.9 & \citenum{Moruzzi1995_ch3oh}\tabularnewline
 34-35 & 4 &  & $E$\textsuperscript{\emph{a}} & 1391.4 & (0.0) &                                     1413.9 &  & 1391.0 &  &                                                                  1396.2 & \citenum{Fehrensen2003JCP}\tabularnewline
 36-37 & 1 & 6$_1$,7$_1$,11$_1$ & $E$\textsuperscript{\emph{b}} & 1428.4 & (0.3) &                    &  &  &  &  1414.0                                                                                           & \citenum{Serrallach1974JMS} \tabularnewline
 38 & 0 & 5$_1$ & $A_1$ & 1444.5 & (0.2) &                                 1526.6 & 1450.2 & 1446.8 &  &                                                              1453.3 & \citenum{Temsamani2003JMS} \tabularnewline
 39-40 & 0 & 5$_1$ & $E$ & 1454.8 & (0.2) &                                1534.7 & 1457.8 & 1455.5 &  &  1454.5\textsuperscript{\emph{d}}                                                                   & \citenum{Serrallach1974JMS} \tabularnewline
 41-42 & 0 & 4$_1$,10$_1$ & $E$ & 1462.5 & (0.3) &                         1619.6 & 1460.3 & 1470.9 &  &                                                                             1473.9
 & \citenum{Temsamani2003JMS} \tabularnewline
 43 & 0 & 4$_1$,10$_1$ & $A_2$ & 1467.9 & (0.3) &                          1631.5 & 1465.0 & 1475.1 &  &                                                                     1465(3)\textsuperscript{\emph{d}}  &               \citenum{Serrallach1974JMS}\tabularnewline
 44-45 & 1,2 & 7$_1$,11$_1$ & $E$\textsuperscript{\emph{c}} & 1471.5 & (0.3) &                          &  &  &  &                                                                                         &     \tabularnewline
 46-47 & 0 & 4$_1$,10$_1$ & $E$ & 1477.4 & (0.3) &                         1631.1 & 1478.7 & 1481.1 &  &                                                                            1477.2\textsuperscript{\emph{d}}  & \citenum{Serrallach1974JMS}\tabularnewline
 48 & 0 & 4$_1$,10$_1$ & $A_1$ & 1481.3 & (0.3) &                          1642.2 & 1484.0 & 1483.8 &  &                                                                     1479.5\textsuperscript{\emph{d}}  & \citenum{Serrallach1974JMS}\tabularnewline
 49 & 1 & 6$_1$ & $A_2$ & 1531.3 & (0.2) &                                  &  &  &  &                                                                                 & \tabularnewline
 50 & 2 & 6$_1$,7$_1$,11$_1$ & $A_1$ & 1540.5 & (0.2) &                     &  &  &  &                                                                                             & \tabularnewline
  \hline\hline                                          
\end{tabular}
     }
\end{center}     
\begin{flushleft}
$^a$~1391.376, 1391.391
$^b$~1428.408, 1428.421
$^c$~1471.546, 1471.557
$^d$~{Ref. \citenum{Temsamani2003JMS} also reports experimental values: 1462.1, 1481.5, 1483.3, 1486.1 in \cm\ assignable to our no. 39-40, 43, 46-47, and 48, energy levels, respectively.}
\end{flushleft}
\end{table*}

 \begin{table*}[!htbp]
  \caption{%
    Vibrational states of CH$_3$OH... [Table \ref{tbl:b=8_assign1} continued.]
  }\label{tbl:b=8_assign2}
  \scalebox{0.9}{
 \begin{tabular}{@{}ccccc c cc cccc@{}}
 \hline\hline
 \# & $\nu_{\tau}$ & SAMs & irrep & $\tilde{\nu}$ &  $\delta$ &  Ref.~\citenum{Fehrensen2003JCP} & Ref.~\citenum{Sibert2005JCP} & Ref.~\citenum{Bowman2007JPCA} & Ref.~\citenum{Lauvergnat2014SA_methanol,Nauts2018MP_methanol}  & \multicolumn{2}{c}{$\tilde{\nu}_\text{Expt.}$} \tabularnewline
\hline
 51-52 & 2 & 8$_1$ & $E$ & 1543.3 & (0.3) &                                 &  &  &  &                                                                                 1548.1 & \citenum{Moruzzi1995_ch3oh}\tabularnewline
 53-54 & 1,2 & 6$_1$,7$_1$,11$_1$ & $E$ & 1577.4 & (0.3) &                    &  &  &  &                                                                                               & \tabularnewline
 55-56 & 2 & 5$_1$,7$_1$,11$_1$ & $E$ & 1656.1 & (0.3) &                    &  &  &  &                                                                                               & \tabularnewline
 57-58 & 1 & 5$_1$ & $E$ & 1662.5 & (0.3) &                                 &  &  &  &                                                                                  & \tabularnewline
 59 & 1 & 4$_1$, 10$_1$ & $A_1$ & 1667.8 & (0.3) &                          &  &  &  &                                                                                         & \tabularnewline
 60 & 1 & 4$_1$, 10$_1$ & $A_2$ & 1668.5 & (0.3) &                          &  &  &  &                                                                                         & \tabularnewline
 61-62 & 1 & 4$_1$, 10$_1$ & $E$ & 1676.4 & (0.3) &                         &  &  &  &                                                                                          & \tabularnewline
 63 & 1,2 & 6$_1$,7$_1$,11$_1$ & $A_2$ & 1711.9 & (0.3) &                     &  &  &  &                                                                                              & \tabularnewline
 64-65 & 2 & 6$_1$ & $E$ & 1731.6 & (0.2) &                                 &  &  &  &                                                                                  & \tabularnewline
 66 & 2 & 6$_1$,7$_1$,11$_1$ & $A_1$ & 1745.1 & (0.3) &                                  &  &  &  &                                                                                 & \tabularnewline
 67 & 1 & 5$_1$ & $A_2$ & 1747.9 & (0.2) &                                  &  &  &  &                                                                                 & \tabularnewline
 68-69 & 1 & 4$_1$, 10$_1$ & $E$ & 1755.7 & (0.3) &                         &  &  &  &                                                                                          & \tabularnewline
 70-71 & 3 & 8$_1$ & $E$ & 1782.0 & (0.3) &                                 &  &  &  &                                                                                  & \tabularnewline
 72-73 & 5 &  & $E$ & 1794.7 & (0.0) &                                      &  &  &  &                                                                             & \tabularnewline
 74 & 2 & 5$_1$ & $A_1$ & 1809.5 & (0.2) &                                  &  &  &  &                                                                                 & \tabularnewline
 75-76 & 2 & 4$_1$, 10$_1$ & $E$ & 1818.9 & (0.3) &                         &  &  &  &                                                                                          & \tabularnewline
 77 & 3 & 7$_1$, 11$_1$ & $A_2$ & 1907.1 & (0.3) &        &  &  &  &                & \tabularnewline
 78 & 3 & 7$_1$, 11$_1$ & $A_1$ & 1907.3 & (0.3) &        &  &  &  &                & \tabularnewline
 79-80 & 2 & 5$_1$ & $E$ & 1941.8 & (0.3) &                                 &  &  &  &                                                                                  & \tabularnewline
 81-82 & 2 & 4$_1$, 10$_1$ & $E$ & 1958.6 & (0.3) &                          &  &  &  &                                                                                         & \tabularnewline
 83-84 & 2 & 5$_1$ & $E$ & 1974.5 & (0.3) &                                 &  &  &  &                                                                                  & \tabularnewline
 85 & 2 & 4$_1$, 10$_1$ & $A_2$ & 1983.4 & (0.3) &     &  &  &  &                                     & \tabularnewline
 86 & 2 & 4$_1$, 10$_1$ & $A_1$ & 1983.5 & (0.3) &     &  &  &  &                                     & \tabularnewline
 87-88 & 3 & 6$_1$ & $E$ & 1986.3 & (0.3) &                                 &  &  &  &                                                                                  & \tabularnewline
 89 & 0 & 8$_2$ & $A_1$ & 2059.4 & (1.5) &                                 2129.6 & 2049.7 & 2038.7 &  &                                                               2055.0 & \citenum{Lees2002PRA}\tabularnewline
 90-91 & 0 & 8$_2$ & $E$ & 2068.8 & (1.5) &                                2138.2 & 2057.5 & 2051.6 &  &                                                                       2063.6 & \citenum{Lees2002PRA}\tabularnewline
 92 & 3 & 8$_1$ & $A_2$ & 2076.8 & (0.3) &                                  &  &  &  &                                                                                 & \tabularnewline
 93 & 4 & 8$_1$ & $A_1$ & 2077.5 & (0.3) &                                  &  &  &  &                                                                                 & \tabularnewline
 94 & 0 & [6$_1$+8$_1$]  & $A_1$ & 2090.7 & (1.2) &                        2194.7 & 2095.3 & 2097.6 &  &                                                                       2100.2 & \citenum{Lees2002PRA}\tabularnewline
 95-96 & 0 & [7$_1$+8$_1$, 11$_1$+8$_1$]\textsuperscript{\emph{a}} & $E$ & 2095.3 & (1.2) &           2186.1 & 2099.9 & 2104.0 &  &                                                                                          2104.9 & \citenum{Lees2002PRA}\tabularnewline
 97 & 0 & [6$_1$+11$_1$, 6$_1$+7$_1$]\textsuperscript{\emph{b}} & $A_1$ & 2122.7 & (1.2) &           2253.4 & 2141.4 & 2152.5 &  &                                                                                    2145.1 & \citenum{Lees2002PRA}\tabularnewline
 98-99 & 0 & [6$_1$+11$_1$, 6$_1$+7$_1$]  & $E$ & 2124.9 & (1.2) &        2245.4& 2143.4 & 2158.2 &  &                                                                                             2147.4 & \citenum{Lees2002PRA}\tabularnewline
 100 & 0 & 11$_1$+8$_1$, 7$_1$+8$_1$ & $E$ & 2181.9 & (1.2) &               &  &  &  &                                                                                                   2182.9 & \citenum{Lees2002PRA}\tabularnewline
 \hline\hline                                              
     \end{tabular}
     }
\begin{flushleft}
$^a$~{Assignment of Ref. \citenum{Lees2002PRA} is $7_1+8_1$.}
$^b$~{Assignment of Ref. \citenum{Lees2002PRA} is $7_2$.}   
\end{flushleft}
     
\end{table*}

%
%
\subsubsection{Comparison with experiment}\label{subsec:compare_exp}
Tables \ref{tbl:b=8_assign1} and \ref{tbl:b=8_assign2} list all computed vibrational energies up to 2182~\cm\ beyond the ZPVE using $b=8$ in the GENIUSH-Smolyak program and the PES13 of CH$_3$OH. 
Based on the analysis shown in the Sec.~S3 of Supporting Information, we decided to use the following basis and grid parameters in the largest computation,  $n_{\tau}=33$ and $M_{\tau}=54$ for the LAM basis and grid size; and 
$b=8$ and $H=22$ for the SAM basis truncation and grid pruning parameters, respectively.
The convergence of the vibrational energies was estimated by comparing the $b=7$ and $b=8$ results. All vibrational energies differ by less than 0.5~\cm\ (indicating an excellent convergence) up to 2000~\cm\ beyond the zero-point vibrational energy. These states can be assigned to the ground or the vibrational fundamental, $n_i=0,1\ (i=1,\ldots,11)$, for each SAM degree of freedom. At the same time, states with up to the $n_\tau=5$th state (4th overtone) of the (low-energy and floppy) torsional mode are observed. 
The current $b=7$ and $b=8$ SAM basis truncation parameters (and other basis and grid details specified earlier) are excellent for the complete vibrational description of this energy range (up to 2000~\cm). 

Beyond 2000~\cm , we observe overtones and combination bands of the SAMs, and correspondingly a larger, up to 1.5~\cm, deviation of the $b=7$ and 8 vibrational energy values is found. For a sub-1~\cm\ convergence, larger $b$ values would be required to confirm the current $b=8$ values or to improve upon them.

The doubly-degenerate pairs of states, dictated by the $C_{3\text{v}}(\text{M})$ molecular symmetry group, are obtained numerically. With the numerical symmetry-adaptation approach during the curvilinear normal coordinate definition (to eliminate numerical noise from the GF calculations), it was possible to converge (with the present basis sizes) the degeneracies better than 0.01 \cm\ with the exception of three pairs of states explicitly listed in the footnotes of the tables.

We also note that the high-energy vibrational fundamentals corresponding to the $\nu_1,\nu_2,\nu_3,$ and $\nu_9$ normal modes (Table \ref{tbl:harm_freq_aveL}), do not appear in the energy range of our computation. For them, it would be necessary to compute more than 100 vibrational states. 

Our results are in good agreement with the experiments, and the deviation is only a few \cm\ for most of the states. 
The energetic ordering in a dense region (12th-16th states) does not agree with experiment, but the computed energies are close to the experimental values. We find that the no. 43, 46-47, 48 computed energies differ
by more than a few \cm\ from the experimental values (Tables~4 and 7 of Ref.~\citenum{Temsamani2003JMS}, and Table~I of Ref.~\citenum{Serrallach1974JMS}).
Interestingly, for these states (no. 43, 46-47, 48), 
our computation agrees better with the older experiment by Serrallach \emph{et al.} in 1974 \cite{Serrallach1974JMS} than with the more recent experiment in 2003 \cite{Temsamani2003JMS}. When the experimental assignments differ from ours, we indicate the experimental assignments in the footnote.

To the best of our knowledge, computed vibrational states in the 1500--2000~\cm\ range are first reported in the present work. The torsional excited states are observed in this region. 

Regarding the strongly mixed states, \#94--99,  the earlier vibrational perturbation theory \cite{Sibert2005JCP} and MULTIMODE reaction path \cite{Bowman2007JPCA} computations use different PESs from ours (PES13) and report values closer to the experiment.  
Regarding our variational vibrational computation, we think that the vibrational energies are converged (within a few~\cm\ for these energy levels, Figure S1 of Supporting Information) with respect to the vibrational basis set size, and the integration grid is also adequately chosen. Therefore, the larger deviation from the experiment (10-23 \cm, \#94--99 in Table~\ref{tbl:b=8_assign2}) most likely indicates imperfections of the PES for these levels. Most importantly, the basis sets in the electronic structure computation, CCSD(T)-F12b/aug-cc-pVDZ, or the fitting error should be considered in future revisions.

\section{Summary, conclusion, and outlook}\label{sec:conclusion}
Full-dimensional (12D) vibrational computations are performed for the methanol molecule using the GENIUSH-Smolyak computer program and the full-dimensional \emph{ab initio} potential energy surface developed by Qu and Bowman~\cite{Qu2013MP_CH3+OH}. 
All vibrational states converged better than 0.5~\cm\ up to 2000~\cm\ beyond the zero-point vibrational energy (ZPVE). 
The ZPVE is estimated to be converged better than 0.05~\cm. 
Automated computation of curvilinear normal coordinates for the small-amplitude vibrations are implemented in the GENIUSH program (in Fortran) in order to minimize the coupling of the small amplitude vibrations necessary for using efficient basis and grid truncation techniques.
During the numerical implementation and computation of the coordinate coefficients, special attention is devoted to suppressing the numerical noise. This ensures strict symmetry adaptation for good convergence of symmetry features according to the $C_{3\text{v}}(\text{M})$ molecular symmetry group in finite basis representations of this floppy molecule. 

The computed and experimental torsional vibrational band origins (where the experimental value is available) are in excellent agreement, within 5~\cm. Most of the non-torsional vibrational band origins are also in good agreement. 
Since the vibrational computation is well converged, this result indicates the good quality of the PES, while some small imperfections, especially at higher energies are identified.
Further development of the PES for the higher-energy range and extensive comparison (perhaps triggering re-assignments) of the experiments would be welcome in the future.

As to further computations and developments on the quantum dynamics side, we plan to compute rovibrational states, by adapting an efficient body-fixed frame to be able to use the vibrational eigenstates computed in this work as a good basis for the rovibrational computation. This development will open the route towards a direct (line-by-line) comparison with high-resolution experiments and the development of an \emph{ab initio} quantum dynamics database for methanol.
In addition, further applications are planned using the path-following GF method and the GENIUSH-Smolyak approach for systems with large amplitude motion(s).

\begin{acknowledgements}
We thank David Lauvergnat for discussions about path-following $L$ coefficients and double groups. 
A. S. thanks the European Union’s Horizon 2022 research and innovation programme under the Marie Skłodowska-Curie Grant Agreement No. 101105452 (QDMAP).
We thank the financial support of the Hungarian National Research, Development, and Innovation Office (FK 142869).
\end{acknowledgements}

%

\clearpage

\input{ch3oh_final_for_arXiv_som_inc}

\end{document}

%% file: ch3oh_final_for_arXiv_som_inc.tex
\setcounter{section}{0}
\renewcommand{\thesection}{S\arabic{section}}
\setcounter{subsection}{0}
\renewcommand{\thesubsection}{S\arabic{section}.\arabic{subsection}}

\setcounter{equation}{0}
\renewcommand{\theequation}{S\arabic{equation}}

\setcounter{table}{0}
\renewcommand{\thetable}{S\arabic{table}}

\setcounter{figure}{0}
\renewcommand{\thefigure}{S\arabic{figure}}

~\\[0.cm]
\begin{center}
\begin{minipage}{0.8\linewidth}
\centering
\textbf{Supplementary Information to } \\[0.25cm]

\textbf{%
Variational Vibrational States of Methanol (12D) 
}
\end{minipage}
~\\[0.5cm]
\begin{minipage}{0.6\linewidth}
\centering

Ayaki Sunaga,$^1$ Gustavo Avila,$^1$ and Edit M\'atyus$^{1,\ast}$ \\[0.15cm]

$^1$~\emph{ELTE, Eötvös Loránd University, Institute of Chemistry, 
Pázmány Péter sétány 1/A, Budapest, H-1117, Hungary} \\[0.15cm]
$^\ast$ edit.matyus@ttk.elte.hu \\
\end{minipage}
~\\[0.15cm]
(Dated: \today)
\end{center}

\vspace{1.cm}
\noindent %
S1. Symmetry features of the basis functions \\
S2. Harmonic frequencies \\
S3. Convergence of the vibrational energies with respect to the basis size \\
S4. Effect of the pruning parameter and transformation matrix \\

\vspace{0.5cm}
\noindent %
The vibrational eigenvectors are uploaded to the Zenodo repository \cite{zenodo:data_ch3oh}.

\clearpage
%
%
\section{Symmetry features of the basis functions}\label{subsec:assign}
The harmonic states computed within the ave-\Ltau\ GF procedure
(Table 3 of the main text) are assigned based on the  \Ctv\ MS group \cite{Bunker2006_molsym}.
The curvilinear normal coordinates are already symmetry adapted within \Ctv, unlike the primitive internal coordinates.
In particular, 
\bea\label{eq:q_xi}
q_{2} &=& \frac{1}{\sqrt{6}}\left(2\Delta r_{\mathrm{CH_{1}}}-\Delta r_{\mathrm{CH_{2}}}-\Delta r_{\mathrm{CH_{3}}}\right), \\ \nonumber
q_{9} &=& \frac{1}{\sqrt{2}}\left(\Delta r_{\mathrm{CH_{2}}}-\Delta r_{\mathrm{CH_{3}}}\right), \\ \nonumber
q_{3} &=& \frac{1}{\sqrt{3}}\left(\Delta r_{\mathrm{CH_{1}}}+\Delta r_{\mathrm{CH_{2}}}+\Delta r_{\mathrm{CH_{3}}}\right)\; .
\eea
The symmetry properties of the \{$q_2, q_9, q_3$\} coordinate set can be determined based on the transformation properties of the coordinates, \emph{e.g.,} the $\bos{X}_{\mathrm{H}_i}$ Cartesian coordinates, the $r_{\text{CH}_i}$ bond length, the $\theta_{\mathrm{H}_i\mathrm{CO}}$ bond angle, or the dihedral angles ($\tau_i$). 
For example, 
\bea\label{eq:12_b-c}
q_9 &=& \Delta r_{\mathrm{CH_2}}-\Delta r_{\mathrm{CH_3}} \\ \nonumber
(12)^* q_9&=& \Delta r_{\mathrm{CH_1}}-\Delta r_{\mathrm{CH_3} } \\ \nonumber
(123) q_9&=& \Delta r_{\mathrm{CH_3}}-\Delta r_{\mathrm{CH_1} } \; .
\eea
The symmetry properties are summarized in Table~\ref{tbl:H_r_theta_tau} (in this work, $\bos{X}_{\mathrm{H}_i}$ was not used, but we have included it for completeness).
Our $q_i$ coordinates are the linear combinations of the $\Delta \bos{\xi} = \bos{\xi}-\bos{\xi}^\mathrm{ref}$ internal displacement coordinates (Eq.~(34) of the main text), 
and thus, the transformation properties of $\bos{\xi}$ and $\bos{\xi}^\mathrm{ref}$ can be determined using Table~\ref{tbl:H_r_theta_tau}.

%
%
\begin{table}[ht]
\caption{$C_{3\text{v}}(\text{M})$ symmetry properties of representations spanned by the coordinate representations of the hydrogen nuclei in the \ce{CH3} unit of methanol: $\bos{X}_{\mathrm{H}_i}$ Cartesian coordinates, $r_{\mathrm{CH}_i}$ distances, $\theta_{\mathrm{H}_i\mathrm{CO}}$ angles, and $\tau_i$  dihedral angles (HO-CH$_i$, where $i=1,2,3$). 
Transformation with respect to the $E^*$ space inversion is also shown, and the $\Gamma$ irreducible decomposition of the $i=1,2,3$ representation is also given. 
}\label{tbl:H_r_theta_tau}
\begin{tabular}{ccccc}
\hline 
\hline 
  & $\bos{X}_{\mathrm{H}_i}$ & $r_{\mathrm{CH}_i}$ & $\theta_{\mathrm{H}_i\mathrm{CO}}$ & $\tau_i$\tabularnewline
\hline 
$E^{*}$ & $-\bos{X}_{\mathrm{H}_i}$ & $r_{\mathrm{CH}_i}$ & $\theta_{\mathrm{H}_i\mathrm{CO}}$ & $-\tau_i$ $^a$ 
\tabularnewline
$\Gamma$ & $a_{2}\oplus e$ & $a_{1}\oplus e$ & $a_{1}\oplus e$ & $a_{2}\oplus e$\tabularnewline
\hline 
\hline 
\end{tabular}
\begin{flushleft}
$^a$~$E^*(\tau_i)=-\tau_i$ corresponds to $\tau_i \rightarrow (2\pi-\tau_i) $ for $ \tau_i \in[0,2\pi]$.
\end{flushleft}
\end{table}

The symmetry properties of the basis functions as a function of the coordinates can be determined based on Table~\ref{tbl:H_r_theta_tau} and considering properties of the function itself.
The SAMs are described with harmonic oscillator basis functions, which can be either an even or odd function of its variable, which is also considered for its transformation under the MS group operations. 
The LAM is described with Fourier functions, and we need to consider that the $2\pi/3$ periodicity of $\tau$ correspond to exchanges of the methyl protons, $i=1,2,3$. 
%
The resulting symmetry features of the Fourier basis functions are summarized in Table~\ref{tbl:sin_cos}. 
When $n\neq3m$ with a non-negative integer $m$, the behavior of the $\sin(n\tau)$ and $\cos(n\tau)$ correspond to a linear transformation of $q_9$ and $q_2$, respectively, Eq.~\eqref{eq:q_xi}, which are assigned to the $e$ irrep. When $n=3m$, both $\sin(n\tau)$ and $\cos(n\tau)$ correspond to $q_3$. 
According to Table~\ref{tbl:sin_cos}, the assignment for the $n=3m$ case is obtained according to
\bea
a_1:\quad(12)^*\cos(\tau)&=&\cos(-\tau)=\cos(\tau) \\ \nonumber
a_2:\quad(12)^*\sin(\tau)&=&\sin(-\tau)=-\sin(\tau)
\eea
Similar relations hold for (23)$^*$ and (31)$^*$. $m$ corresponds to the number of nodes in the region of $(0+2k\pi/3,2\pi/3+2k\pi/3)$ with $k=0,1,2$. 

\begin{table}[ht]
\caption{%
Values and symmetry assignment of the Fourier basis functions at $\tau=\pi/3,\pi,5\pi/3$. $n=3m+k$ with a non-negative integer $m$ and $k=0,1,2$. 
}\label{tbl:sin_cos}
\begin{tabular}{cccccc}
\hline\hline
 & $\tau$ & $\pi/3$ & $\pi$ & $5\pi/3$& $\Gamma$ \tabularnewline
 \hline
$n\neq3m$ & $\sin(n\tau)$ & $\sqrt{3}/2$ & 0 & $-\sqrt{3}/2$& $e$ \tabularnewline
 & $\cos(n\tau)$ & $\pm1/2$ & $\mp1$ & $\pm1/2$& $e$ \tabularnewline
$n=3m$ & $\sin(n\tau)$ & 0 & 0 & 0& $a_2$ \tabularnewline
 & $\cos(n\tau)$ & $\pm1$ & $\pm1$ & $\pm1$& $a_1$ \tabularnewline
 \hline\hline
\end{tabular}
\end{table}

\section{Harmonic frequencies}
Table \ref{tbl:harm_freq} shows that our implementation code provides the harmonic frequencies at the three minima equivalently within 0.005~\cm. 
The numerical precision of the Fortran-based implementation is checked and confirmed by comparison with an implementation using the Wolfram Mathematica symbolic algebra program \cite{Mathematica} (Eqs.~(4)--(11) of Ref.~\citenum{Daria2022JMS_HCOOH}). 
The double-precision implementation of the PES subroutine limits the precision of the computation, but Mathematica can provide numerically more stable Hessians, than our double-precision Fortran code, which improves the numerical equivalence of the $\tau = 60^\circ,180^\circ,300^\circ$ structures. 
The harmonic frequencies of the GF method obtained using the Fortran code (Sec.~2.3 of the main text) agree with Mathematica within 0.01~\cm.

%
%
\begin{table}[!htbp]
\caption{%
Harmonic frequencies (\cm) at the three equilibrium minima computed with the Wolfram Mathematica and the Fortran GF implementations and PES13~\cite{Qu2013MP_CH3+OH}. $\tau=60^\circ,180^\circ,300^\circ$. $\delta_{\tau} = \nu_{60}-\nu_{\tau}$ with $\tau=180^\circ,300^\circ$. `0.0' means the absolute value of $\delta_{\tau}$ is smaller than $2\times10^{-11}$. The corresponding state label, description, the dominant internal coordinates, and symmetry label in the $C_\text{s}$ point group are also provided.}
\label{tbl:harm_freq}
\scalebox{1.0}{
\begin{tabular}{@{}lll|rrr|rrr@{}}
\hline\hline
  & & & \multicolumn{3}{c|}{Fortran} & \multicolumn{3}{c}{Mathematica}\tabularnewline
description & coord & sym &60$^\circ$ & $\delta_{180^\circ}$ & $\delta_{300^\circ}$ & 60$^\circ$ & $\delta_{180^\circ}$ & $\delta_{300^\circ}$ \tabularnewline
\hline
 $\nu$(\ce{OH}) & $r_\mathrm{OH}$ & $a'$ &3869.4223 & 7$\times 10^{-6}$ & 4$\times 10^{-5}$ & 3869.4223 & 3$\times 10^{-8 }$& 0.0\tabularnewline
 $\nu$(\ce{CH3})$_{\mathrm{asym}}$  & $r_\mathrm{CH}$ &$a'$ &3150.2661 & 7$\times 10^{-6}$ & 3$\times 10^{-5 }$& 3150.2657 & 2$\times 10^{-6 }$& 0.0\tabularnewline
 $\nu$(\ce{CH3})$_{\mathrm{asym}}$ & $r_\mathrm{CH}$&$a''$ & 3068.8035 & 4$\times 10^{-5}$ & 2$\times 10^{-4}$ & 3068.8031 & $-$3$\times 10^{-7 }$& 0.0\tabularnewline
 $\nu$(\ce{CH3})$_{\mathrm{sym}}$& $r_\mathrm{CH}$ &  $a'$ & 3016.1881 & $-$6$\times 10^{-6}$ & $-$3$\times 10^{-5}$ & 3016.1879 & 1$\times 10^{-6}$ & 0.0\tabularnewline
 $\delta$(\ce{CH3})$_{\mathrm{asym}}$ & $\varphi_2$ &  $a'$ & 1527.1540 & $-$2$\times 10^{-6}$ & $-$3$\times 10^{-3}$ & 1527.1553 & 1$\times 10^{-5}$ & 0.0 \tabularnewline
$\delta$(\ce{CH3})$_{\mathrm{asym}}$ & $\varphi_1$ &$a''$ &1498.9571 & 4$\times 10^{-5}$ & $-$3$\times 10^{-3}$ & 1498.9627 & 3$\times 10^{-6 }$& 0.0\tabularnewline
 $\delta$(\ce{CH3})$_{\mathrm{sym}}$ & $\theta_\mathrm{HCO}$ &  $a'$ & 1475.9578 & $-$2$\times 10^{-5}$ & 9$\times 10^{-5}$ & 1475.9577 & $-$1$\times 10^{-5 }$& 0.0\tabularnewline
 $\delta$(\ce{COH}) & $\theta_\mathrm{COH}$ &  $a'$ & 1376.7803 & $-$7$\times 10^{-5}$ & $-$1$\times 10^{-3 }$& 1376.7807 & 3$\times 10^{-5 }$& 0.0\tabularnewline
$\rho$(\ce{CH3})$_{\mathrm{}}$& $\theta_\mathrm{HCO}$ &  $a''$ &1196.2107 & 6$\times 10^{-5}$ & $-$1$\times 10^{-3}$ & 1196.2125 & $-$2$\times 10^{-5}$ & 0.0\tabularnewline
$\rho$(\ce{CH3})$_{\mathrm{}}$ &$\theta_\mathrm{HCO}$ & $a'$ &1067.0133 & $-$5$\times 10^{-6}$ & $-$1$\times 10^{-3}$ & 1067.0131 & $-$3$\times 10^{-6}$ & 0.0\tabularnewline
 $\nu$(\ce{CO})$_{\mathrm{}}$&  $r_\mathrm{CO}$ & $a'$ &1065.1486 & $-$1$\times 10^{-5}$ & 4$\times 10^{-6}$ & 1065.1485 & 2$\times 10^{-7 }$& 0.0\tabularnewline
 tor(OH) &  $\tau$ & $a''$ & 286.6917 & $-$2$\times 10^{-5}$ & 8$\times 10^{-6}$ & 286.6913 & 6$\times 10^{-7 }$& 0.0 \tabularnewline
\hline\hline
\end{tabular}
}
\end{table}

\section{Convergence of the vibrational energies with respect to the basis size}\label{subsec:res:converge}
The basis and grid sizes have been extensively tested. 
Regarding the $\tau$ torsional degree of freedom, the 1D-torsional computations (Table~\ref{tbl:1D_grid}) provided guidance. At least $n_{\tau}=33$ torsional basis functions were necessary to converge the lowest 100 torsional states up to ca. 2500~\cm\ beyond the ZPVE. 
The corresponding number of integration points was determined (Table~\ref{tbl:12D_grid}) to be 
$M_{\tau}=54$ for a 0.01~\cm\ convergence. 

\begin{figure}[ht]
\includegraphics[width=0.5\textwidth]{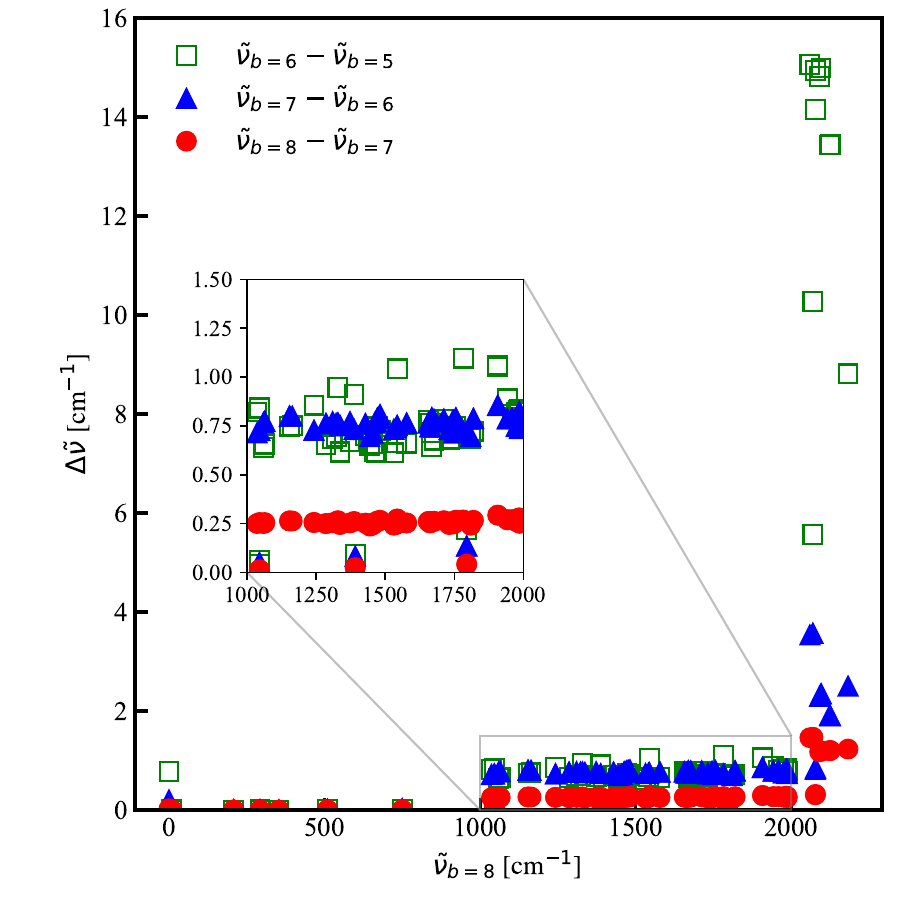}
\caption{\label{fig:b5678} 
Convergence of the vibrational energy, measured from the ZPVE, with respect to the basis set truncation parameter, $b$, in 12D vibrational computations of CH$_3$OH with GENIUSH-Smolyak and PES13. 
For the numerical integration, $H=14+b$ was used. 
The ZPVEs for $b=5,6,7$, and 8 are $\tilde{v}_0=11109.020, 11108.248, 11108.047$, and $11108.023\ \mathrm{~cm}^{-1}$, respectively.
}
\end{figure}

Regarding the SAMs, we have carried out computations with increasingly large $b$ values, (Eqs.~41 and 42 of the main text), $b=5,6,7,$ and 8. 
A systematic convergence is observed by increasing the basis set size, and the differences between $\tilde{\nu}_{b=8}$ and $\tilde{\nu}_{b=7}$ are less than 1 \cm\ for the most states (Figure~\ref{fig:b5678}). 

\begin{figure}[ht]
 \includegraphics[width=0.5\textwidth]{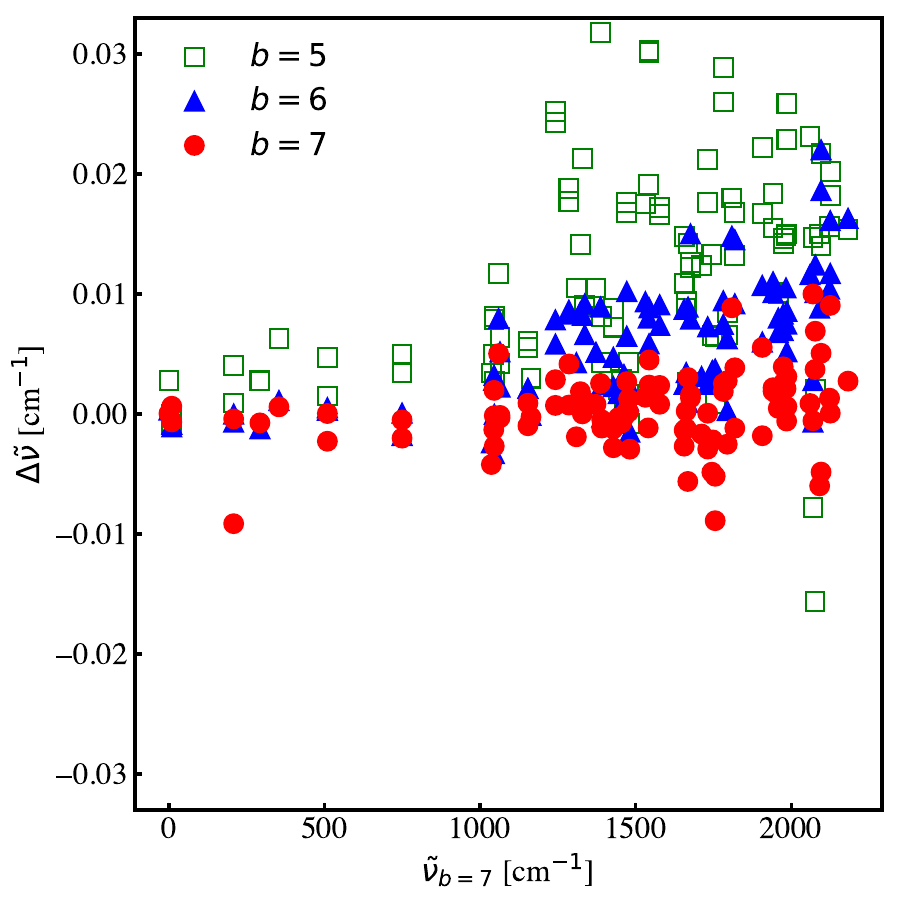}
  \caption{%
    \label{fig:aveL-aveL_tau} 
    Comparison of the ave-\Ltau\ and ave-$L$ vibrational energies,
    $\Delta\tilde{\nu}=\tilde{\nu}_{\mathrm{ave}\mathchar`-L(\tau)}-\tilde{\nu}_{\mathrm{ave}\mathchar`-L}$, 
    for the $b=5,6,$ and 7 basis and the $H=14+b$ grid sizes, obtained with the GENIUSH-Smolyak program and the PES13. 
    The vibrational energies are referenced to their respective ZPVEs; ave-$L$ with $b=5,6,7$: $\tilde{v}_0=11109.020,	11108.248$, $11108.047 \mathrm{~cm}^{-1}$; and ave-\Ltau\ with $b=5,6,7$: $\tilde{v}_0=11109.023, 11108.249$, $11108.047 \mathrm{~cm}^{-1}$.
  }
\end{figure}

The difference between ave-\Ltau\ and ave-$L$ in the 12D computation using several basis set sizes is shown in Figure~\ref{fig:aveL-aveL_tau}. It is observed that by increasing the basis size, the different coordinate choices have a smaller role. The difference of the vibrational energies decreases with an increasing basis size, and in the infinite basis limit, all (correct) coordinate definitions are, of course, expected to give the same energies.
For $b=7$, the largest absolute deviation of the ave-$L$ and ave-\Ltau\ energies was 0.01~\cm.

 The ZPVE obtained using ave-\Ltau\ is \emph{larger} by a few $10^{-3}$ \cm\ than the ZPVE obtained with ave-$L$ (same basis size, $b=5$ and 6). This ZPVE difference is tiny, but we add that we would have expected the other way round: a lower value from ave-\Ltau, which is expected to provide a better coordinate representation (corresponding to the numerical scheme used to construct the coordinate definition), so this deviation is probably due to numerical noise. Regarding 1D model computations (Table~3 in the main text), the relaxed path model, sym-\xitau, performed better. 
Regarding ave-\Ltau, it would be necessary to rule out any (remaining) numerical noise, but the convergence level with $b=7$ is already sufficient for our purposes. In the large-scale production computations, we will use the sym-\xitau\ and ave-$L$ combination of the coordinate definition.

%
%
\begin{table}
\caption{%
Vibrational energy levels of CH$_3$OH, in \cm, referenced to the ZPVE (\#1) using a 1D vibrational model defined with sym-\xitau\ and ave-$\xi$ reference structures and using GENIUSH and the PES13 for an increasing number of DVR grid points. 
The highest-energy levels that can be converged with the DVR basis and grid are highlighted in bold.
}
\label{tbl:1D_grid}
\begin{tabular}{@{}crrrrrr@{}}
\hline\hline
  & 27$^\ast$ & 29$^\ast$ & 31$^\ast$ & 33$^\ast$ & 35$^\ast$   & 35$^\ast$ \tabularnewline
 \#  & sym-\xitau & sym-\xitau & sym-\xitau  & sym-\xitau  & sym-\xitau   &  ave-$\xi$  \tabularnewline
 \hline
1 & 120.465 & 120.466 & 120.466 & 120.465 & 120.466  & 235.155 \tabularnewline
2 & 9.909 & 9.909 & 9.909 & 9.909 & 9.909  & 13.558\rule[0mm]{0mm}{4mm}\tabularnewline
3 & 9.909 & 9.910 & 9.910 & 9.909 & 9.910  & 13.558 \tabularnewline
4 & 199.462 & 199.461 & 199.461 & 199.462 & 199.461  & 168.864\rule[0mm]{0mm}{4mm}\tabularnewline
5 & 199.462 & 199.461 & 199.462 & 199.462 & 199.461  & 168.864 \tabularnewline
6 & 288.260 & 288.261 & 288.261 & 288.262 & 288.261  & 271.206\rule[0mm]{0mm}{4mm}\tabularnewline
7 & 342.401 & 342.398 & 342.398 & 342.399 & 342.398  & 305.663\rule[0mm]{0mm}{3mm}\tabularnewline
8 & 502.659 & 502.656 & 502.658 & 502.659 & 502.658  & 475.820\rule[0mm]{0mm}{4mm}\tabularnewline
9 & 502.659 & 502.661 & 502.658 & 502.659 & 502.659  & 475.820 \tabularnewline
10 & 745.859 & 745.857 & 745.853 & 745.858 & 745.857  & 719.407\rule[0mm]{0mm}{4mm}\tabularnewline
11 & 745.859 & 745.858 & 745.861 & 745.858 & 745.857  & 719.407 \tabularnewline
12 & 1046.138 & 1046.143 & 1046.143 & 1046.138 & 1046.143  & 1018.897\rule[0mm]{0mm}{4mm}\tabularnewline
13 & 1046.448 & 1046.440 & 1046.440 & 1046.447 & 1046.440  & 1019.021\rule[0mm]{0mm}{3mm}\tabularnewline
14 & 1402.635 & 1402.626 & 1402.633 & 1402.634 & 1402.626  &1373.709\rule[0mm]{0mm}{4mm}\tabularnewline
15 & 1402.635 & 1402.640 & 1402.633 & 1402.634 & 1402.640  & 1373.709 \tabularnewline
16 & \textbf{1814.413} & 1814.411 & 1814.402 & 1814.412 & 1814.411  & 1783.393\rule[0mm]{0mm}{4mm}\tabularnewline
17 & \textbf{1814.413} & 1814.411 & 1814.420 & 1814.412 & 1814.411  & 1783.393 \tabularnewline
18 & 2281.336 & \textbf{2281.427} & 2281.428 & 2281.419 & 2281.428  & 2247.897\rule[0mm]{0mm}{4mm}\tabularnewline
19 & 2281.521 & \textbf{2281.431 }& 2281.431 & 2281.444 & 2281.431  & 2247.897\rule[0mm]{0mm}{3mm}\tabularnewline
20 & 2803.582 & 2803.535 & \textbf{2803.593 }& 2803.596 & 2803.581  & 2767.167\rule[0mm]{0mm}{4mm}\tabularnewline
21 & 2803.582 & 2803.646 &\textbf{ 2803.593} & 2803.596 & 2803.607  & 2767.167 \tabularnewline
22 & 3378.663 & 3380.834 & 3380.822 & \textbf{3380.853} & 3380.851  & 3341.172\rule[0mm]{0mm}{4mm}\tabularnewline
23 & 3378.663 & 3380.845 & 3380.876 & \textbf{3380.853} & 3380.851  & 3341.172 \tabularnewline
24 & 3922.872 & 4011.090 & 4013.128 & 4013.134 & 4013.136  &  3969.894\rule[0mm]{0mm}{4mm}\tabularnewline
25 & 4112.207 & 4011.163 & 4013.198 & 4013.211 & 4013.206  &  3969.894\tabularnewline
\hline\hline
\end{tabular}
\begin{flushleft}
$^\ast$~Number of Fourier DVR points and functions.
\end{flushleft}
\end{table}

\begin{table}
\caption{%
Convergence test regarding the number of $M_{\tau}$ integration grid points for the torsional degree of freedom in a 12D computation of CH$_3$OH with the GENIUSH-Smolyak program and the PES13. The vibrational energy levels are, in \cm, referenced to the ZPVE (\#1). $b = 4, H = 18, N_{\tau}=32$, sym-\xitau\ and ave-$L$ are used.  $\delta_{M_{\tau}} = \nu_{M_{\tau} + 2}-\nu_{M_{\tau}}$, in \cm.
The largest deviations (in an absolute value) are highlighted in bold.
}
\label{tbl:12D_grid}
\begin{tabular}{@{}c rrrrr@{}}
\hline\hline
$M_{\tau}$& $\delta_{46}$ & $\delta_{48}$ & $\delta_{52}$ & $\delta_{54}$ & 56\tabularnewline
\hline
1  & $ 8\times 10^{-3 }$ & $ -7\times 10^{-3 }$ & $ 5\times 10^{-4 }$ & $ -2\times 10^{-3 }$ & 11109.9395\tabularnewline
2  & $ -1\times 10^{-2 }$ & $ 1\times 10^{-2 }$ & $ -2\times 10^{-4 }$ & $ 4\times 10^{-3 }$ & 9.131\rule[0mm]{0mm}{4mm}\tabularnewline
3 & $ -2\times 10^{-2 }$ & $ 8\times 10^{-3 }$ & $ 1\times 10^{-3 }$ & $ 1\times 10^{-3 }$ & 9.132\tabularnewline
4  & $ -9\times 10^{-3 }$ & $ 3\times 10^{-3 }$ & $ 4\times 10^{-3 }$ & $ 4\times 10^{-3 }$ & 207.973\rule[0mm]{0mm}{4mm}\tabularnewline
5  & $ -4\times 10^{-3 }$ & $ 1\times 10^{-2 }$ & $ -5\times 10^{-3 }$ & $\bos{9\times 10^{-3}}$ & 207.982\tabularnewline
6  & $ -4\times 10^{-3 }$ & $ 8\times 10^{-3 }$ & $ 3\times 10^{-4 }$ & $ -5\times 10^{-4 }$ & 292.461\rule[0mm]{0mm}{4mm}\tabularnewline
7  & $ -8\times 10^{-3 }$ & $ 7\times 10^{-3 }$ & $ 5\times 10^{-3 }$ & $ -2\times 10^{-3 }$ & 353.570\rule[0mm]{0mm}{3mm}\tabularnewline
8  & $ -1\times 10^{-2 }$ & $ 1\times 10^{-2 }$ & $ 6\times 10^{-4 }$ & $ 7\times 10^{-4  }$& 509.460\rule[0mm]{0mm}{4mm}\tabularnewline
9  & $ -2\times 10^{-2 }$ & $ 9\times 10^{-3 }$ & $ 2\times 10^{-4 }$ & $ 3\times 10^{-3  }$& 509.462\tabularnewline
10  & $ 5\times 10^{-3 }$ & $ -5\times 10^{-4 }$ & $ 2\times 10^{-3 }$ & $ -6\times 10^{-4 }$ & 749.719\rule[0mm]{0mm}{4mm}\tabularnewline
11 & $ -2\times 10^{-2 }$ & $ 7\times 10^{-3 }$ & $ -4\times 10^{-3 }$ & $ -4\times 10^{-3 }$ & 749.720\tabularnewline
12  & $ -1\times 10^{-2 }$ & $ 7\times 10^{-3 }$ & $ 2\times 10^{-3 }$ & $ 9\times 10^{-4  }$& 1044.713\rule[0mm]{0mm}{4mm}\tabularnewline
13& $ 2\times 10^{-2 }$ & $ -1\times 10^{-2 }$ & $ -3\times 10^{-3 }$ & $ 3\times 10^{-3  }$& 1045.117\rule[0mm]{0mm}{3mm}\tabularnewline
14  & $ 5\times 10^{-3 }$ & $ -7\times 10^{-3 }$ & $ 4\times 10^{-3 }$ & $ -6\times 10^{-3 }$ & 1049.921\rule[0mm]{0mm}{3mm}\tabularnewline
15  & $ -2\times 10^{-2 }$ & $ 8\times 10^{-3 }$ & $ 3\times 10^{-3 }$ & $ 3\times 10^{-3 }$ & 1058.930\rule[0mm]{0mm}{4mm}\tabularnewline
16  & $ -1\times 10^{1 }$ & $ 2\times 10^{-3 }$ & $ 3\times 10^{-3 }$ & $ 4\times 10^{-3 }$ & 1058.935\tabularnewline
17  & $ -5\times 10^{0 }$ & $\bos{ 2\times 10^{-2 }}$ & $ \bos{1\times 10^{-2}}$ & $ -8\times 10^{-3  }$& 1071.675\rule[0mm]{0mm}{4mm}\tabularnewline
18  & $\bos{ -9\times 10^{1 }}$ & $ 1\times 10^{-2 }$ & $ 2\times 10^{-3 }$ & $ 4\times 10^{-3 }$ & 1076.314\rule[0mm]{0mm}{4mm}\tabularnewline
19  & $\bos{-1\times 10^{2 }}$ & $ 1\times 10^{-2 }$ & $ -2\times 10^{-3 }$ & $ 4\times 10^{-3 }$ & 1076.315\tabularnewline
20 & $ -9\times 10^{1 }$ & $ 2\times 10^{-2 }$ & $ -5\times 10^{-3 }$ & $ 3\times 10^{-3  }$& 1166.609\rule[0mm]{0mm}{4mm}\tabularnewline
\hline\hline
\end{tabular}
\end{table}

%
%

%
%

%
%
\section{Effect of the pruning parameter and transformation matrix}
The numerical values of the vibrational energy levels measured from the zero-point vibrational energy (ZPVE) are listed in Tables \ref{tbl:b_5678_1-50} and \ref{tbl:b_5678_51-100}. The main text defines the pruning parameter for the basis sets ($b$) and the transformation matrix (ave-$L$ and ave-\Ltau). The order of the 89th and 90-91th energy levels at $b=5$ is exchanged following the symmetry assignment of $b=8$.

\begin{table*}[ht]
\caption{
Vibrational states of CH$_3$OH (12D) computed with the GENIUSH-Smolyak program and the PES13 potential energy surface of Ref.~\cite{Qu2013MP_CH3+OH}. 
 Vibrational energies, $\tilde{\nu}$ in \cm, referenced to the ZPVE (\#1) and assignments obtained using the $b=5,6,7$ and 8 basis sets. Regarding the coordinate definition, sym-\xitau\ is used in all computations, and the coefficient vector, ave-$L$ or ave-\Ltau, is specified in the column headings. 
 SAMs: assignment of the curvilinear normal modes $1_n, 2_n, ..., 11_n$ $(n=0,1,\ldots)$ in Table~3 of the main text, $n=0$ (zero) excitation is not shown.  `$[\left.\ldots\right]$' labels the largest contribution(s) from strongly mixed states. 
 Sym: irrep label of $C_{3\text{v}}(\text{M})$ according to Table~3 of the main text and corresponding discussion.
 The $E$ state degeneracy is (numerically) converged better than 0.1~\cm\ for all $b$ values. 
} 
\label{tbl:b_5678_1-50}
\begin{tabular}{cccc|rrrr|rrr}
\hline
& & & & \multicolumn{4}{c}{ave-$L$} & \multicolumn{3}{|c}{ave-\Ltau}\tabularnewline
$\#$ & $\nu_{\tau}$ & SAMs & sym   & 5 & 6 & 7 & 8 & 5 & 6 & 7\tabularnewline
\hline 1 & 0 & & $A_1$ &                            11109.020 & 11108.248 & 11108.047 & 11108.023 & 11109.023 & 11108.249 & 11108.047\tabularnewline
2-3 & 0 & & $E$ &                                 9.121 & 9.117 & 9.117 & 9.118 & 9.121 & 9.116 & 9.118\tabularnewline
4-5 & 1 & & $E$ &                                 208.0 & 208.0 & 208.0 & 208.0 & 208.0 & 208.0 & 208.0\tabularnewline
6 & 1 & & $A_2$ &                                   292.4 & 292.4 & 292.4 & 292.4 & 292.4 & 292.4 & 292.4\tabularnewline
7 & 2 & & $A_1$ &                                   353.5 & 353.6 & 353.6 & 353.6 & 353.6 & 353.6 & 353.6\tabularnewline
8-9 & 2 & & $E$ &                                 509.4 & 509.3 & 509.3 & 509.3 & 509.4 & 509.3 & 509.3\tabularnewline
10-11 & 3 & & $E$ &                              749.5 & 749.5 & 749.5 & 749.5 & 749.5 & 749.5 & 749.5\tabularnewline
12 & 0 & $8_1$ & $A_1$ &                           1038.1 & 1037.2 & 1036.5 & 1036.3 & 1038.1 & 1037.2 & 1036.5\tabularnewline
13 & 3 & & $A_2$ &                                 1044.3 & 1044.3 & 1044.2 & 1044.2 & 1044.3 & 1044.3 & 1044.2\tabularnewline
14 & 4 & & $A_1$ &                                 1045.1 & 1045.1 & 1045.0 & 1045.0 & 1045.1 & 1045.1 & 1045.0\tabularnewline
15-16 & 0 & $8_1$ & $E$ &                        1047.4 & 1046.6 & 1045.9 & 1045.6 & 1047.4 & 1046.6 & 1045.9\tabularnewline
17 & 0 & $6_1, 7_1, 11_1$ & $A_1$ &                1061.1 & 1060.4 & 1059.7 & 1059.4 & 1061.1 & 1060.4 & 1059.7\tabularnewline
18-19 & 0 & $6_1, 7_1, 11_1$ & $E$ &             1065.7 & 1065.0 & 1064.2 & 1064.0 & 1065.7 & 1065.0 & 1064.2\tabularnewline
20-21 & 0 & $11_1, 10_1$ & $E$ &                 1155.6 & 1154.8 & 1154.0 & 1153.8 & 1155.6 & 1154.8 & 1154.0\tabularnewline
22 & 0 & $11_1, 10_1$ & $A_2$ &                    1165.3 & 1164.6 & 1163.8 & 1163.5 & 1165.3 & 1164.6 & 1163.8\tabularnewline
23-24 & 1 & $8_1$ & $E$ &                        1244.0 & 1243.1 & 1242.4 & 1242.2 & 1244.0 & 1243.1 & 1242.4\tabularnewline
25-26 & 1 & $6_1, 7_1, 11_1$ & $E$ &             1287.2 & 1286.6 & 1285.8 & 1285.5 & 1287.2 & 1286.6 & 1285.8\tabularnewline
27 & 1 & $6_1, 7_1, 11_1$ & $A_1$ &                1311.2 & 1310.6 & 1309.8 & 1309.5 & 1311.3 & 1310.6 & 1309.8\tabularnewline
28 & 1 & $6_1, 7_1, 11_1$ & $A_2$ &                1324.2 & 1323.5 & 1322.7 & 1322.5 & 1324.2 & 1323.5 & 1322.7\tabularnewline
29 & 1 & $8_1$ & $A_2$ &                           1330.6 & 1329.7 & 1328.9 & 1328.6 & 1330.6 & 1329.7 & 1328.9\tabularnewline
30-31 & 0 & $6_1$ & $E$ &                        1338.1 & 1337.4 & 1336.7 & 1336.4 & 1338.1 & 1337.4 & 1336.7\tabularnewline
32 & 1,2 & $6_1, 7_1, 11_1$ & $A_1$ &                1374.3 & 1373.6 & 1372.8 & 1372.6 & 1374.3 & 1373.6 & 1372.8\tabularnewline
33 & 2 & $8_1$ & $A_1$ &                           1389.1 & 1388.2 & 1387.5 & 1387.2 & 1389.2 & 1388.2 & 1387.5\tabularnewline
34-35 & 4 & & $E$ &                    1391.6 & 1391.5 & 1391.4 & 1391.4 & 1391.6 & 1391.5 & 1391.4\tabularnewline
36-37 & 1 & $6_1, 7_1, 11_1$ & $E$ &   1430.1 & 1429.4 & 1428.7 & 1428.4 & 1430.1 & 1429.4 & 1428.7\tabularnewline
38 & 0 & $5_1$ & $A_1$ &                           1446.1 & 1445.4 & 1444.7 & 1444.5 & 1446.1 & 1445.4 & 1444.7\tabularnewline
39-40 & 0 & $5_1$ & $E$ &                        1456.5 & 1455.8 & 1455.1 & 1454.8 & 1456.5 & 1455.8 & 1455.1\tabularnewline
41-42 & 0 & $4_1, 10_1$ & $E$ &                  1464.1 & 1463.5 & 1462.7 & 1462.5 & 1464.1 & 1463.5 & 1462.7\tabularnewline
43 & 0 & $4_1, 10_1$ & $A_2$ &                     1469.5 & 1468.9 & 1468.1 & 1467.9 & 1469.5 & 1468.9 & 1468.1\tabularnewline
44-45 & 1,2 & $7_1, 11_1$ & $E$ &        1473.4 & 1472.6 & 1471.8 & 1471.6 & 1473.4 & 1472.6 & 1471.8\tabularnewline
46-47 & 0 & $4_1, 10_1$ & $E$ &                  1479.1 & 1478.4 & 1477.6 & 1477.4 & 1479.2 & 1478.4 & 1477.7\tabularnewline
48 & 0 & $4_1, 10_1$ & $A_1$ &                     1483.1 & 1482.4 & 1481.6 & 1481.3 & 1483.1 & 1482.4 & 1481.6\tabularnewline
49 & 1 & $6_1$ & $A_2$ &                           1532.9 & 1532.3 & 1531.6 & 1531.3 & 1532.9 & 1532.3 & 1531.6\tabularnewline
50 & 2 & $6_1, 7_1, 11_1$ & $A_1$ &                1542.1 & 1541.4 & 1540.7 & 1540.5 & 1542.1 & 1541.5 & 1540.7\tabularnewline
\hline
\end{tabular}
\end{table*}

\begin{table*}[ht]
\caption{
Vibrational states of $\mathrm{CH}_3 \mathrm{OH} \ldots$ [Table \ref{tbl:b_5678_1-50} continued.]
} 
\label{tbl:b_5678_51-100}
\begin{tabular}{cccc|rrrr|rrr}
\hline
& & & & \multicolumn{4}{c}{ave-$L$} & \multicolumn{3}{|c}{ave-\Ltau}\tabularnewline
$\#$ & $\nu_{\tau}$ & SAMs & sym   & 5 & 6 & 7 & 8 & 5 & 6 & 7\tabularnewline
\hline 51-52 & 2 & $8_1$ & $E$ &                                      1545.3 & 1544.3 & 1543.5 & 1543.3 & 1545.4 & 1544.3 & 1543.5\tabularnewline
53-54 & 1 & $6_1, 7_1, 11_1$ & $E$ &                                  1579.1 & 1578.5 & 1577.7 & 1577.4 & 1579.1 & 1578.5 & 1577.7\tabularnewline
55-56 & 1 & $5_1, 7_1, 11_1$ & $E$ &                                  1657.9 & 1657.1 & 1656.3 & 1656.1 & 1657.9 & 1657.1 & 1656.3\tabularnewline
57-58 & 1 & $5_1$ & $E$ &                                             1664.3 & 1663.5 & 1662.8 & 1662.5 & 1664.3 & 1663.5 & 1662.8\tabularnewline
59 & 1 & $4_1, 1_1$ & $A_1$ &                                           1669.4 & 1668.8 & 1668.0 & 1667.8 & 1669.4 & 1668.8 & 1668.0\tabularnewline
60 & 1 & $4_1, 10_1$ & $A_2$ &                                          1670.2 & 1669.5 & 1668.7 & 1668.5 & 1670.2 & 1669.5 & 1668.7\tabularnewline
61-62 & 1 & $4_1, 10_1$ & $E$ &                                       1678.1 & 1677.4 & 1676.6 & 1676.4 & 1678.1 & 1677.4 & 1676.6\tabularnewline
63 & 2 & $6_1, 7_1, 11_1$ & $A_2$ &                                     1713.7 & 1713.0 & 1712.2 & 1711.9 & 1713.7 & 1713.0 & 1712.2\tabularnewline
64-65 & 2 & $6_1$ & $E$ &                                             1733.3 & 1732.6 & 1731.9 & 1731.6 & 1733.3 & 1732.6 & 1731.9\tabularnewline
66 & 2 & $6_1$ & $A_1$ &                                                1746.9 & 1746.1 & 1745.3 & 1745.1 & 1746.9 & 1746.1 & 1745.3\tabularnewline
67 & 1 & $5_1$ & $A_2$ &                                                1749.5 & 1748.8 & 1748.1 & 1747.9 & 1749.5 & 1748.8 & 1748.1\tabularnewline
68-69 & 1 & $4_1, 10_1$ & $E$ &                                       1757.5 & 1756.8 & 1756.0 & 1755.7 & 1757.6 & 1756.8 & 1756.0\tabularnewline
70-71 & 3 & $8_1$ & $E$ &                                             1784.1 & 1783.0 & 1782.3 & 1782.0 & 1784.1 & 1783.0 & 1782.3\tabularnewline
72-73 & 5 & & $E$ &                                                   1795.1 & 1794.9 & 1794.7 & 1794.7 & 1795.1 & 1794.9 & 1794.7\tabularnewline
74 & 2 & $5_1$ & $A_1$ &                                                1811.1 & 1810.4 & 1809.7 & 1809.5 & 1811.1 & 1810.4 & 1809.7\tabularnewline
75-76 & 2 & $4_1, 10_1$ & $E$ &                                       1820.7 & 1820.0 & 1819.2 & 1818.9 & 1820.7 & 1820.0 & 1819.2\tabularnewline
77 & 3 & $7_1, 11_1$ & $A_2$ &     1909.3 & 1908.3 & 1907.4 & 1907.1 & 1909.3 & 1908.3 & 1907.4\tabularnewline
78 & 3 & $7_1, 11_1$ & $A_1$ &     1909.5 & 1908.5 & 1907.6 & 1907.3 & 1909.5 & 1908.5 & 1907.6\tabularnewline
79-80 & 2 & $5_1$ & $E$ &                                             1943.8 & 1942.9 & 1942.1 & 1941.8 & 1943.8 & 1942.9 & 1942.1\tabularnewline
81-81 & 2 & $4_1, 10_1$ & $E$ &                                          1960.5 & 1959.7 & 1958.9 & 1958.6 & 1960.5 & 1959.7 & 1958.9\tabularnewline
83-84 & 2 & $5_1$ & $E$ &                                             1976.4 & 1975.5 & 1974.8 & 1974.5 & 1976.4 & 1975.5 & 1974.8\tabularnewline
85 & 2 & $4_1, 10_1$ & $A_2$ &   1985.4 & 1984.5 & 1983.7 & 1983.4 & 1985.4 & 1984.5 & 1983.7\tabularnewline
86 & 2 & $4_1, 10_1$ & $A_1$ &   1987.4 & 1986.6 & 1985.8 & 1985.5 & 1987.5 & 1986.6 & 1985.8\tabularnewline
87-88 & 3 & $6_1$ & $E$ &                                             1988.1 & 1987.3 & 1986.6 & 1986.3 & 1988.1 & 1987.3 & 1986.6\tabularnewline
89 & 0 & $8_2$ & $A_1$ &                                                2084.1 & 2064.3 & 2060.8 & 2059.4 & 2084.1 & 2064.4 & 2060.8\tabularnewline
90-91 & 0 & $8_2$ & $E$ &                                             2079.4 & 2073.9 & 2070.3 & 2068.8 & 2079.4 & 2073.8 & 2070.3\tabularnewline
92 & 3 & $8_1$ & $A_2$ &                                                2092.9 & 2077.9 & 2077.1 & 2076.8 & 2092.8 & 2077.9 & 2077.1\tabularnewline
93 & 4 & $8_1$ & $A_1$ &                                                2092.9 & 2078.7 & 2077.9 & 2077.5 & 2092.9 & 2078.7 & 2077.9\tabularnewline
94 & 0 & {$\left[6_1+8_1\right]$} & $A_1$ &                             2109.0 & 2094.2 & 2091.9 & 2090.7 & 2109.0 & 2094.2 & 2091.9\tabularnewline
95-96 & 0 & {$\left[7_1+8_1, 11_1+8_1\right]$} & $E$ &   2113.8 & 2098.8 & 2096.5 & 2095.3 & 2113.8 & 2098.8 & 2096.5\tabularnewline
97 & 0 & {$\left[6_1+11_1, 6_1+7_1\right]$} & $A_1$ &      2139.2 & 2125.8 & 2123.9 & 2122.7 & 2139.2 & 2125.8 & 2123.9\tabularnewline
98-99 & 0 & {$\left[6_1+11_1, 7_1+11_1\right]$} & $E$ &               2141.4 & 2128.0 & 2126.1 & 2124.9 & 2141.5 & 2128.0 & 2126.1\tabularnewline
100 & 0 & $11_1+8_1, 7_1+8_1$ & $E$ &                                   2194.5 & 2185.7 & 2183.2 & 2181.9 & 2194.5 & 2185.7 & 2183.2\tabularnewline
\hline
\end{tabular}
\end{table*}

\begin{figure}
    \centering
    \includegraphics[width=1.0
    \linewidth]{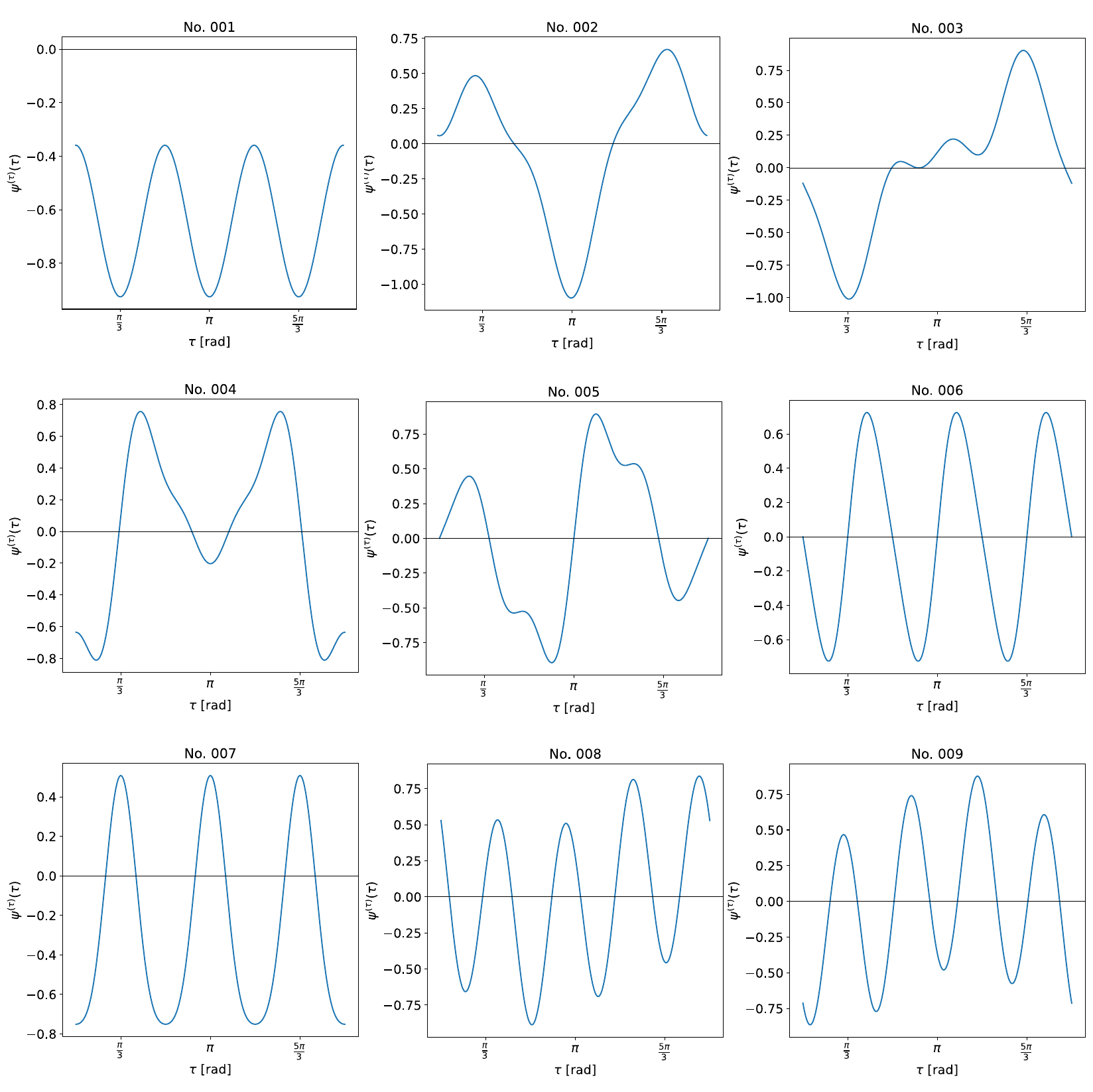}
    \caption{Visualization of the torsional basis functions ($\psi^{\tau}(\tau)$) in the 12D vibrational states listed in Table~6 of the main text up to \#9. Only the dominant configurations (Supporting Information file vib\_b=8.txt) are considered for visualization. The properties of the total wave functions are listed in Table~\ref{tbl:b_5678_1-50} and in Table~6 of the main text.}
    \label{fig:1D_wavefunc}
\end{figure}

The torsional part of the 12D wave function is shown in Figure \ref{fig:1D_wavefunc}. 
The shapes of the wave functions are not exactly symmetric because of the numerical noise and the finite basis size.

%
%
